\documentclass[%
 reprint,
 amsmath,amssymb,
 aps,
]{revtex4-2}

\usepackage{graphicx}% Include figure files
\usepackage{dcolumn}% Align table columns on decimal point
\usepackage{bm}
\usepackage{cancel}
\usepackage{graphics}
\usepackage[normalem]{ulem} 
 \usepackage{color}
	\newcommand{\AM}{\textcolor{black}} %comments Adrian
\begin{document}

%\preprint{}

\title{Quantitative analysis of phase formation and growth in ternary mixtures upon evaporation of one component}


\author{Stela Andrea~Muntean}
\affiliation{Department of Engineering and Physics, Karlstad University, Sweden}
\author{Vi C.E.~Kronberg}
\affiliation{Department of Mathematics and Computer Science, Eindhoven University of Technology, The Netherlands
}
\altaffiliation{Department of Engineering and Physics, Karlstad University, Sweden}
\author{Matteo Colangeli}
\affiliation{Department of Information Engineering, Computer Science and Mathematics,  University of L'Aquila, Italy}
\author{ Adrian Muntean}
\affiliation{Department of Mathematics and Computer Science,  Karlstad University, Sweden}
\author{Jan van Stam}
\affiliation{Department of Engineering and Chemical Sciences, Karlstad University, Sweden}
\author{Ellen Moons}
\affiliation{ Department of Engineering and Physics, Karlstad University, Sweden}
\author{Emilio N.M.~Cirillo}
\affiliation{Department of Basic and Applied Sciences for Engineering (SBAI), Sapienza University of Rome, Italy}

%\thanks{}
%\affiliation{}


\date{\today}

\begin{abstract}We perform a quantitative analysis of Monte--Carlo simulations results of phase separation in ternary blends upon evaporation of one component.
Specifically, we calculate the average domain size and plot it as a function of simulation time to compute the exponent of the obtained power--law.
We compare and discuss results obtained by two different methods, for three different models: 2D binary--state model (Ising model),
2D ternary--state model with and without evaporation.
For the ternary--state models, we study additionally the dependence of the domain growth on concentration, temperature and initial composition.
We reproduce the expected 1/3 exponent for the Ising model, while
for the ternary--state model without evaporation and for the one with evaporation we obtain lower values of the exponent. It turns out that phase separation patterns that can form in this type of systems are complex. The obtained quantitative results give valuable insights towards devising computable theoretical estimations of size effects on morphologies as they occur in the context of organic solar cells.
\end{abstract}

%\keywords{}

\maketitle

%\section{}
%\subsection{}
%\subsubsection{}
\section{Introduction}

Morphology formation is one of the key factors in the processing of multi--component thin films from solution. In applications such as organic solar cells the photoactive layer is a thin film of electron donor and electron acceptor molecules.
The \AM{internal structure of the} morphology of this layer \AM{plays a very important role} for the electric charge generation and collection; cf. e.g. \cite{Fuwen2018}. The photoactive layer is produced within a solution with organic solvent - a distinguishing feature of the organic solar cells compared to the more conventional, silicon-based, photovoltaic systems. %\begin{color}{red} REFERENCES NEEDED\end{color}
Morphologies are formed by phase separation of the electron donor and electron acceptor molecules during the evaporation of the solvent.

\AM{In the framework of this paper, lattice spin systems are used} to understand \AM{size effects on} different morphologies as well as \AM{characteristic time scales} observed in domain growth phenomena.
A paradigmatic model is the widely studied 2D Ising model \cite{Onsager1944}, that served as a model system for the investigation of phase separation in magnetic materials and of novel transport mechanisms under nonequilibrium conditions \cite{Col18}.
For such a model, it has been well established that in the spinodal decomposition regime, namely \AM{for} zero external field and subcritical temperature, the average diameter of the \AM{formed} domains scales with time as a power law,  where the exponents depend on the details of the dynamics.
For nonconservative and conservative dynamics the exponents are $1/2$, and respectively, $1/3$; see e.g. \cite{B1994}.
\AM{Interestingly,} these results \AM{were shown} to be robust with respect to \AM{slight modifications} of the Hamiltonian describing the equilibrium properties of the system;  \AM{we refer the reader for instance to \cite{CGS1997}. On the other hand}, different interesting phenomena are observed when the dynamics are modified \AM{more drastically} by introducing, for example, shear effects \AM{as in} \cite{CGS2005}.

In order to describe \AM{the formation of internal structures} in the presence of an evaporating solvent \AM{as typical to applications in the context of organic solar cells}, a three--state model is needed.
Two straightforward generalisations of the Ising model with nearest neighbour interaction between three--state spin variables are the Blume--Capel \cite{B1966,BEG1971,C1966} and the Potts \cite{P1952} models. \AM{Both these models} have received much attention for their ability to model different physical situations both at equilibrium \cite{Wu1982} and out of equilibrium \cite{CO1996,CN2013,CNS2017,LL2016}.
The distinguishing feature between these two models is that in the Blume--Capel model, interfaces between different spins have different costs.
The dynamics of phase separation for the Potts model \AM{seen} in the spinodal decomposition regime are not completely understood: a rather clear scenario is described for nonconservative dynamics for three and four state spins \cite{SM1995}, whereas the understanding is \AM{only} partial when spin variables with larger cardinality are considered \cite{BFCLP2007}.
To the authors's knowledge, no references are available for the Potts nor the Blume--Capel model with conservative dynamics.

\AM{Three--state lattice models were used in \cite{previous-paper} and \cite{lambda-paper} as an efficient tool to study phase separation patterns in ternary mixtures that allow the evaporation of one of the components  as an alternative to phase--field models used in the same context \cite{Michels,Schaefer1,Harting,Harting2}.} %\begin{color}{red}check order of references!\end{color}
In our previous work  \cite{previous-paper}, we focused our attention on exploring the relevant parameters that influence the morphology formation.
This is a subject of large interest in the community and was discussed in many experimental  and computational studies\cite{Alison,Battaile,Rolf,Lowengrub,Michels,Du,Lyons2,Lyons1,Schaefer1,Schaefer2,Harting,Negi}; see also \cite{Olle} for related work done for stochastic models for competitive growth of phases.
%\begin{color}{red} More experimental references are needed. \end{color}
In the present work, the emphasis falls onto the quantitative analysis of the results for \AM{relevant choices} of parameters.
\AM{Quantitative studies are important particularly when one wants to compare morphologies obtained with different computational models or to compare computational and experimental results.  In general,} different \AM{models and computational} methods give results on different length scales. In some \AM{approaches} the characteristic length scales are \AM{incorporated into} the model parameters, such as the parameters \AM{arising} in the \AM{structure of the} interaction potential \AM{for} atomistic molecular simulations. Even in apparently scale--free models, the nature of the \AM{modeling} assumptions \AM{induces} a length--scale range for the results. Having in view potential applications of our three--state lattice models, it is of a primary concern to identify relevant quantities that capture the essence of domain size evolution across the scales.
%, hence more generally applicable for a range of methods.
The quantitative analysis presented here is an important step towards a more general quantification \AM{and eventual classification} of morphology pictures obtained by different experimental and computational methods.

In this paper, we follow \AM{ up } the ideas developed in our previous work \cite{previous-paper} and give a quantitative study of the phase separated domains \AM{using a slightly simplified version of our original model generalising the Blume--Capel and  Potts models}.
The three possible states of the spin variable are denoted \AM{here} by $-1$, $0$, and $+1$: ``$0$'' is interpreted as an interaction site of solvent molecules, whereas ``$\pm1$''  represent interaction sites of the other two components.
We prefer this more general formulation, since we do not take in account the different molecular weights or volumes of the three components and hence each interaction site occupies the same volume in the lattice.
In case the molecular weights of the three components are comparable, the spin variables could be interpreted as different molecules, but in our case of interest, where the non--evaporating molecules, e.g.~polymers, have different sizes and are much larger than the solvent molecules, this assumption does not hold.

To study the phase separation in the ternary mixture upon evaporation of one of the components, we consider a spin model with a Kawasaki--like dynamics \cite{K1972} \AM{governed} by the Metropolis algorithm \cite{MRRTT1953} to account for energy differences associated to possible spin exchanges and computed using appropriate boundary conditions.
The Kawasaki dynamics is modified here such that evaporation of the $0$ component is allowed.
\AM{Keeping track of the solvent evaporation} is crucial \AM{for} the study of the morphology formation in solution--borne thin films, used e.g.~in the preparation of the active layer in organic photovoltaics.
To implement such a mechanism, the $0$s in the first row of the lattice are removed (evaporation) and replaced by a $+1$ or a $-1$ with probability chosen proportionally to the initial fractions.
The dynamics start with a randomly chosen configuration consisting of a fixed fraction of the three different spin species and evolve until an a priori small concentration of solvent is reached.
The problem we study is related to the spinodal decomposition with the new ingredient of the evaporation of the zero component. We have to keep in mind that our model relies on assumptions and will not correspond to all aspects of the physical reality and that further improvements are possible to make the model more realistic. This falls outside the scope of the present work.

The paper is organised as follows: we first define the model and describe the tools that we rely on to investigate the domain formation under evaporation.
Then, we discuss our numerical results \AM{firstly in the absence of the evaporation, and then, when the evaporation of the solvent, i.e.~of the 0 component, is involved}.
After discussing the effect of the temperature \AM{parameter} on the \AM{overall} system dynamics, a short summary of our findings  concludes  the paper.

\section{Model and methods}
In this section, we firstly define the model and then illustrate the tools that we use to measure the domain growth.
In the last part of this section, we briefly illustrate these methods by discussing the standard Ising case. \AM{For more details on the Monte Carlo method and its variations, we refer the reader e.g. to the monographs \cite{Barkema,Landau-Binder}.}

\subsection{Model}
Let $\Lambda$ be the square $\{1,\dots,L\}^2$ endowed with periodic boundary conditions.
An element of $\Lambda$ is called \emph{site} and two sites are said to be \emph{nearest neighbours} if their Euclidean distance is one.
A pair of nearest neighbouring sites is called a \emph{bond}.
\AM{We} associate the \emph{spin variable} $\sigma(x,y)\in\{-1,0,+1\}$ with each site $(x,y)\in\Lambda$ and define the total \emph{energy} using the \emph{Hamiltonian}
\begin{equation}
\label{mod000}
H(\sigma)
=
\frac{1}{2}
\sum_{(x,y),(x',y')\in\Lambda}
\!\!\!\!
\!\!\!\!
J_{\sigma(x,y)\sigma(x',y')}
,\;\;
J =\begin{bmatrix}
		0 & 1 & 4\\
		1 & 0 & 1\\
		4 & 1 & 0
\end{bmatrix},
\end{equation}
for any configuration $\sigma\in\{-1,0,+1\}^\Lambda$, where the rows and columns of the interaction matrix $J$ refer respectively to $-1$, $0$, and $+1$.
In this case, there is no cost associated with self--interaction (\AM{the main} diagonal of $J$), a relatively small cost between $\pm1$ and $0$ sites, and a relatively large cost of an interface between $-1$ and $+1$.
Such a choice of interaction matrix corresponds to a well--studied parametrisation of the Blume--Capel model with zero magnetic field and zero chemical potential; \AM{see e.g.} \cite{CN2013,CO1996,LL2016}. Note that this \AM{specific} structure of the interaction matrix \AM{$J$} promote\AM{s the }phase separation of the components, irrespective of the presence or absence of the evaporation.

Consider the integer time variable $t\ge0$.
Fix the parameter $\beta>0$ and refer to $1/\beta$ as the \emph{temperature}.
Fix $c_{-1},c_0,c_{+1}\in[0,1)$ with \AM{the constraint} $c_{-1}+c_0+c_{+1}=1$. \AM{They are} the \AM{corresponding} fractions of $-1$, $0$, and $+1$ spins in the initial configuration, \AM{that is} at time $t=0$.

The stochastic evolution $\sigma_t$ is constructed by repeating at each time $t>0$ the following steps $2L^2$ times:
\begin{itemize}
\item[i)]
choose a bond at random with uniform probability;
\item[ii)]
if the bond is of the type $((x,L),(x,1))$ and $\sigma(x,L)=0$, then replace the spin zero at the site $(x,L)$ by $+1$ with probability $c_{+1}/(1-c_0)$ and by $-1$ with probability $1-c_{+1}/(1-c_0)=c_{-1}/(1-c_0)$ (say that the zero \emph{evaporated});
\item[iii)]
otherwise, let $\Delta$ be the difference of energy between the configuration obtained by exchanging the spins at the two sites of the bond and the actual configuration;  exchange the two spins at the sites of the bond with probability $1$ if $\Delta<0$ and \AM{with probability $\exp\{-\beta\Delta\}$ if $\Delta\geq 0$.}
\end{itemize}
\AM{We} stop the dynamics when the total number of zeroes in the system becomes smaller than $L^2/10$. \AM{Based on this description,
the dynamics are of Kawasaki type} complemented with a Metropolis updating rule with the addition of the evaporation rule (i.e. step ii) of the algorithm).

\subsection{Methods}
A common measure of the average domain size is obtained by fixing a cutoff for the \emph{two--point correlation function}.
More precisely, for any $s \in \{s_x,s_y\}$ let
\begin{equation} \label{eq:correlation}
	G(s,t)=
	\frac{1}{L^2}\sum_{(x,y)\in \Lambda}
	\sigma_t(x,y)\sigma_t(x+s_x,y+s_y)
\end{equation}
be the two--point correlation function.
Moreover, we also consider the horizontal and vertical two--point correlation functions $G_x(r,t)=G((r,0),t)$ and $G_y(r,t)=G((0,r),t)$, where $r$ is an integer number.
The direction--dependent two--point correlation functions typically decrease from their maximum value at $s = 0$ in an oscillatory fashion, such that it is possible to estimate the size of the domains by fixing a cut--off and finding the value at which the two correlation function intersect such a cut--off.
In this way, we shall find an estimate for \AM{the horizontal  and vertical diameters of the domains, $R_x$ and $R_y$, respectively}.
%\sout{In the paper the cut--off will be always chosen equal to $0.5$.
%???SIMON: is $0.5$ the value of the cutoff that you used in you computations? \begin{color}{blue} Should be 0.1, but I can't be 100\% certain as the filenames do not have the information (sorry about that)\end{color} ???}

Another well--established domain--size measurement is based on the first momenta of the \emph{structure factor}.
More precisely, for any $(k_x,k_y)$ in the first Brillouin zone $\{-\pi, \ -\pi + 2 \pi/L, \ \dots, \ \pi - 2\pi/L, \ \pi\}^2$, let
\begin{equation} \label{eq:struct}
	C((k_x,k_y),t)
	= \frac {1}{L^2} \bigg| \sum_{(x,y)\in\Lambda} \sigma_t(x,y) \, \text{e}^{i \, (k_x x+k_y y)} \bigg | ^2
\end{equation}
be the structure factor.
Note that the quantity in the absolute value above is simply the Fourier transform of the configuration at Monte Carlo time $t$, and can thus be evaluated using any form of fast--Fourier--transform technique to speed up execution.
Hence, another way of estimating the horizontal and the vertical \AM{diameters} of the domains is by redefining $R_x$ and $R_y$ as follows:
\begin{equation} \label{eq:sF}
	R_{\alpha} = \frac{\sum_{(k_x,k_y)} C((k_x,k_y),t)} {\sum_{(k_x,k_y)} |k_\alpha| C((k_x,k_y),t)} \;,
\end{equation}
where $\alpha \in \{x,y\}$ and each summation is carried out over the first Brillouin zone.

\subsection{Domain growth in the 2D Ising model}
\begin{figure}
\centering
\begin{tabular}{cc}
	\includegraphics[width = 0.22\textwidth]{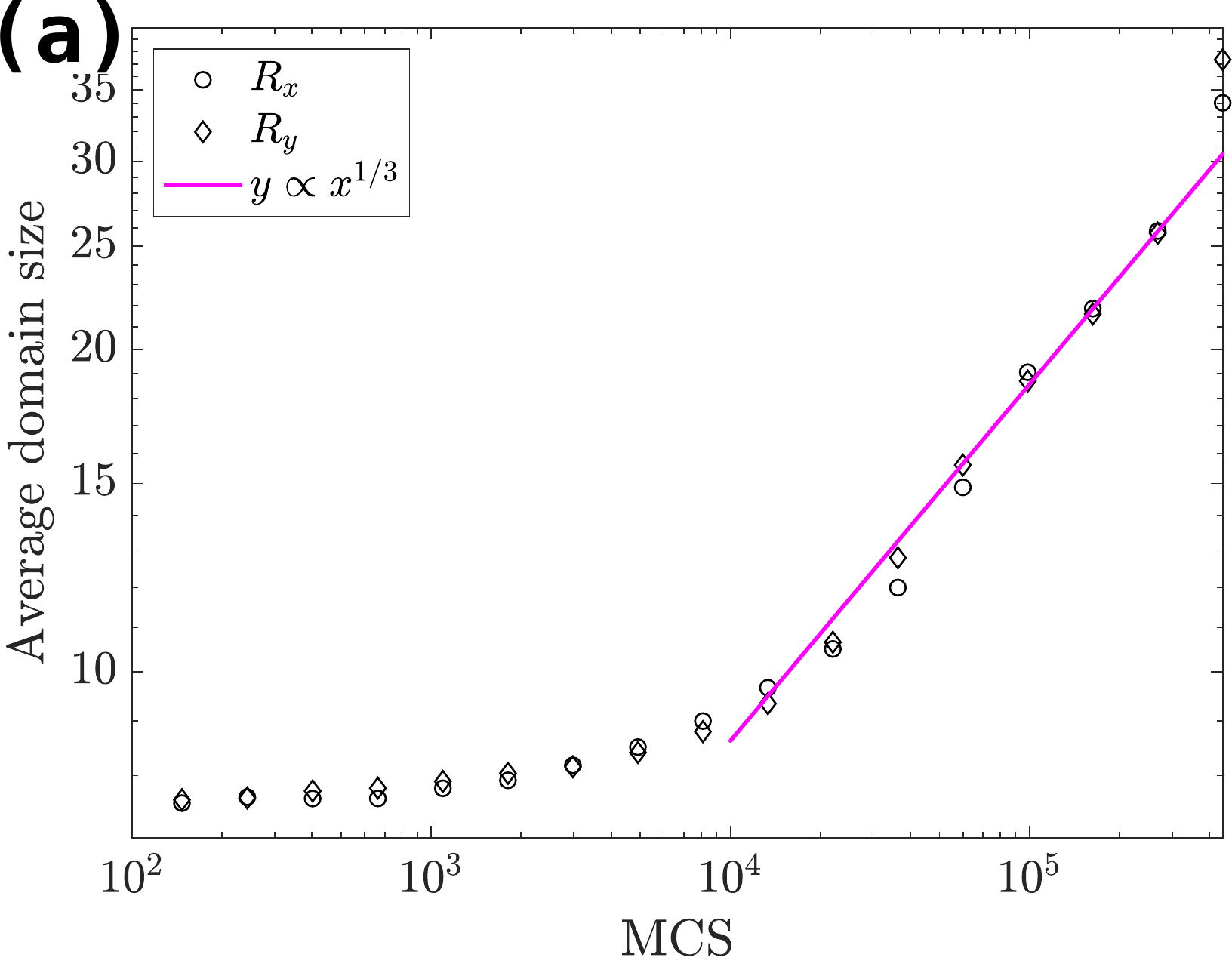}
&
         \includegraphics[width = 0.215\textwidth]{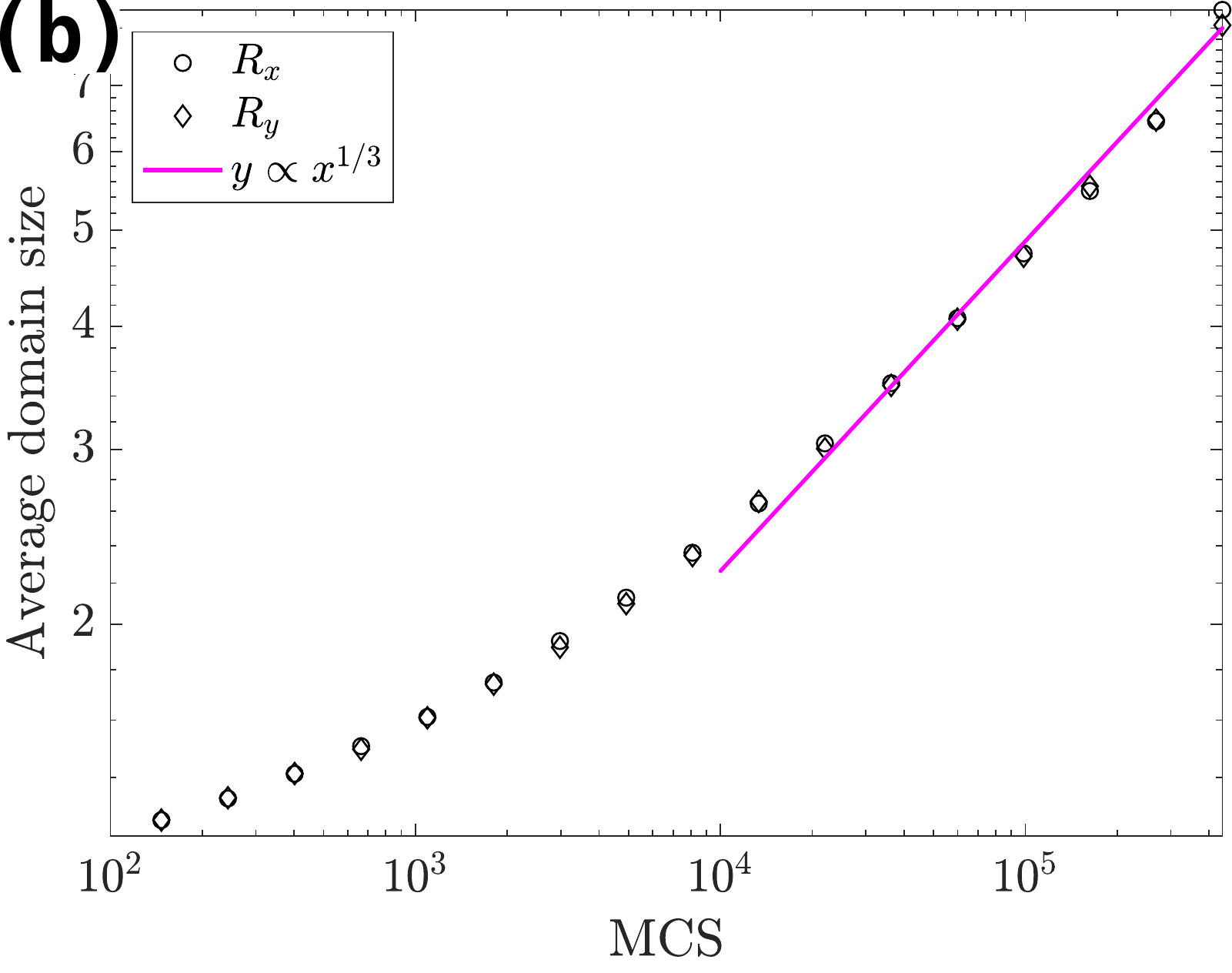}
\end{tabular}
\caption{Single history for the Kawasaki Ising dynamics with $L = 512$ and $\beta = 1.0$. Correlation function measure and structure function measure of the horizontal and vertical domain size in the left and in the right panel, respectively.
}
\label{fig:Kawasaki}
\end{figure}

As a test of this methodology, we consider the classical two--dimensional Ising model under Kawasaki dynamics, for which the growth exponent of $1/3$ has been extensively verified; \AM{we refer the reader for instance to the works} \cite{B1994,CGS1997}.
The results \AM{for our case} are shown in the left and right panel of Fig.~\ref{fig:Kawasaki}.
Here, we report the resulting domain growth of the same run as evaluated using \AM{both} the correlation function and the structure factor.
The previously established $1/3$ exponent is recovered, lending credit to the chosen methodology.

\AM{Based on} Fig.~\ref{fig:Kawasaki}, \AM{we note that the two methods yield quite similar power law structures, i.e. , the same exponent with different constant prefactors}.
This yields a discrepancy in the absolute values of the domain sizes calculated \AM{based on these two} methods.
Since we do not have a physically imposed lengthscale (i.e., no physical meaning of length \AM{is included in our lattice models}), this is not a critical distinction.
It is more important \AM{to observe} how the domains grow, hence \AM{we wish to compare} the exponents \AM{in} the power laws mainly at long times. \AM{Sometimes, also  short timescales could be of interest if power laws of suitable observables are detected.}

Both the correlation function and structure factor measure show that two different regimes can be distinguished: the initial one in which the domains start to be formed by coalescence of equal spins and the second one, characterised by the power law scaling, in which the already formed domains grow in time.
This last regime will be addressed as the \emph{growing} or \emph{scale invariant} regime.

\section{Numerical results}
After having checked the validity of the different methodologies on the Kawasaki Ising dynamics, we now study the three--state model introduced above.
We \AM{explore} the model for different choices of the initial fraction of zeroes $c_0$.
\AM{For} all the simulations, \AM{we set} $c_{-1}=c_{+1}$, so that the initial number of minuses and pluses will be equal.
Due to the evaporation mechanism, the ratio between minuses and pluses will oscillate slightly \AM{during the overall evolution}, while the fraction of zeroes  will progressively decrease.

\subsection{Domain growth in the three state model without evaporation}
\begin{figure}[!ht]
\centering
\begin{tabular}{cc}
\includegraphics[width = 0.22\textwidth]{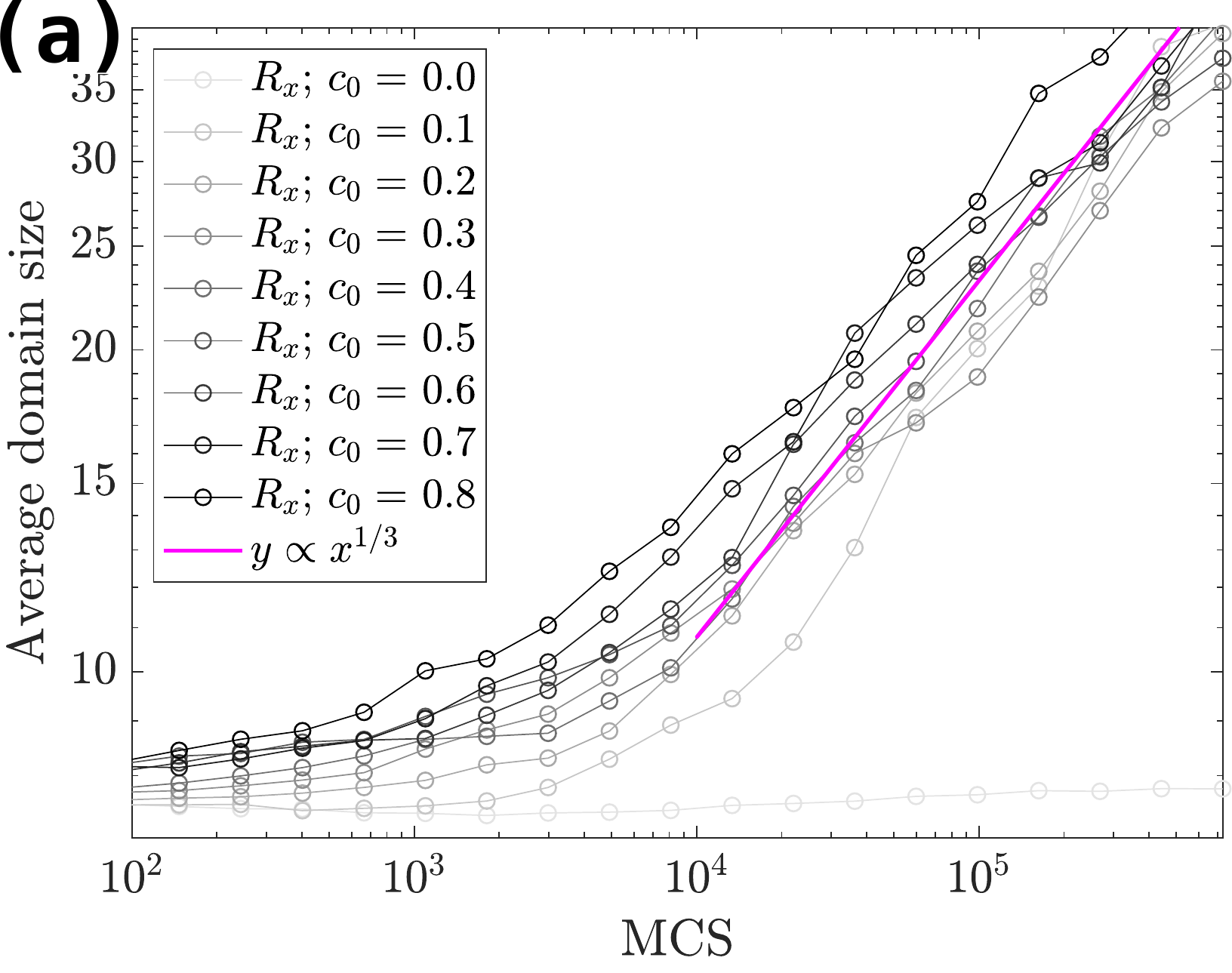}
&
\includegraphics[width = 0.218\textwidth]{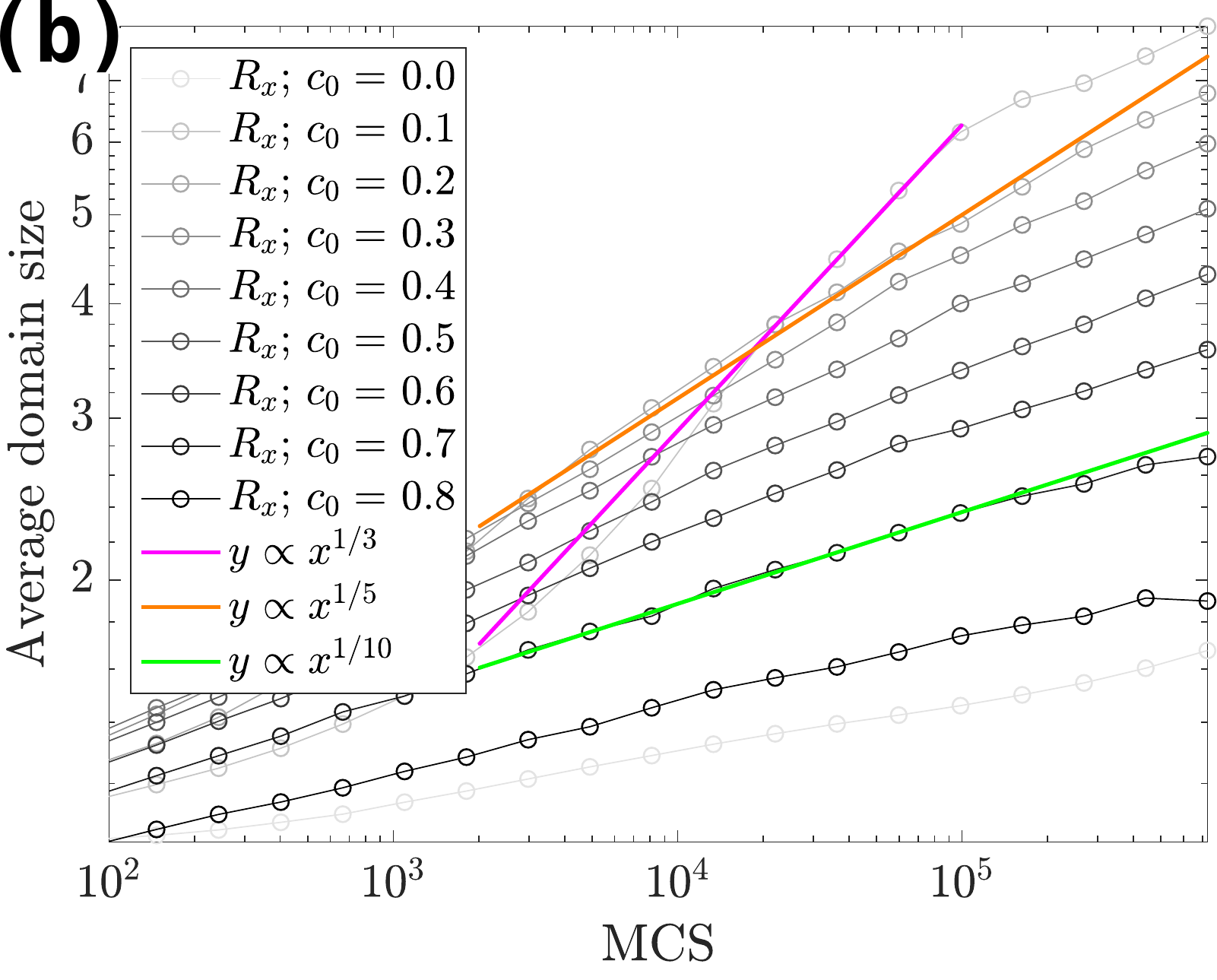}
\end{tabular}
\caption{Three state model with $L=512$ and $\beta=1.0$.
Estimate of the horizontal size of domains via two--point correlation function
on the left and structure factor on the right.
One single system history for each solvent concentration $c_0$ is considered.}
\label{fig:Potts}
\end{figure}

The results for the growth of domains obtained with the two different methods for calculating $R_x$ from the same Monte Carlo simulation of the three state model are shown in Fig.\ \ref{fig:Potts}.
\AM{We omit to show the results for $R_y$ as}  in the absence of evaporation  domains are isotropic, as in the Ising case (see Fig.~\ref{fig:Kawasaki}).
Data for different values of the initial zeroes concentration $c_0$ are reported \AM{here}.
The connecting lines between the data points do not imply piecewise linear regression.
The $x^{1/3}$ line is shown as reference, since a $1/3$ exponent of the power law for the domain size as a function of time is the observed behaviour in the scale invariant regime via the correlation function (left panel). This is consistent with the findings \AM{for a} conserved dynamics \AM{as} reported in \cite{B1994}.
\begin{figure}[!ht]
\centering
\begin{tabular}{cccc}
\includegraphics[width = 0.09\textwidth]{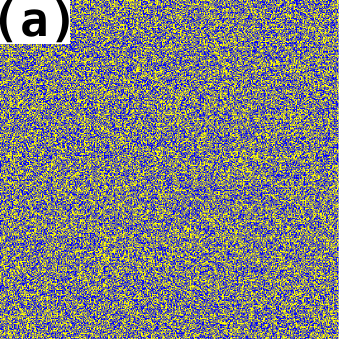} &
\includegraphics[width = 0.09\textwidth]{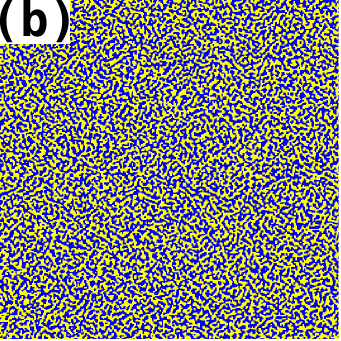} &
\includegraphics[width = 0.09\textwidth]{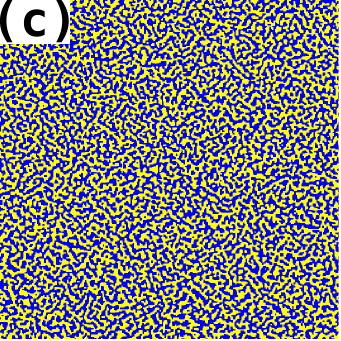} &
\includegraphics[width = 0.09\textwidth]{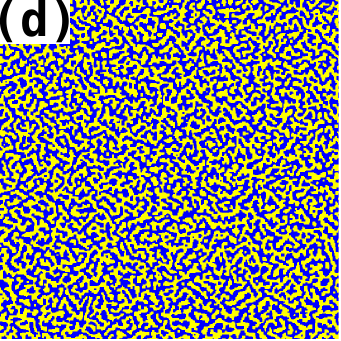}
\\[0.3cm]
\includegraphics[width = 0.09\textwidth]{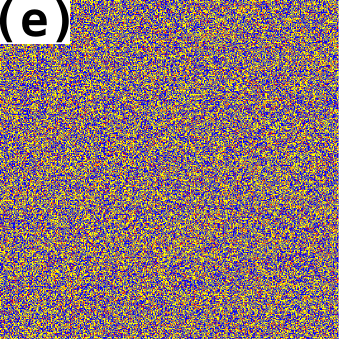} &
\includegraphics[width = 0.09\textwidth]{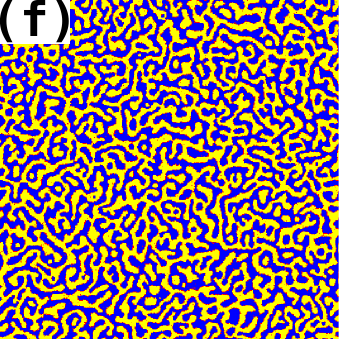} &
\includegraphics[width = 0.09\textwidth]{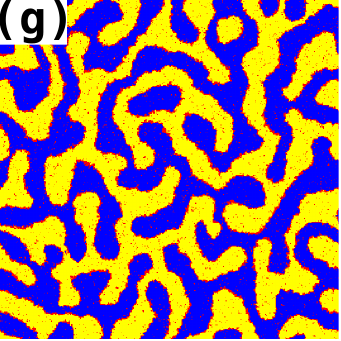} &
\includegraphics[width = 0.09\textwidth]{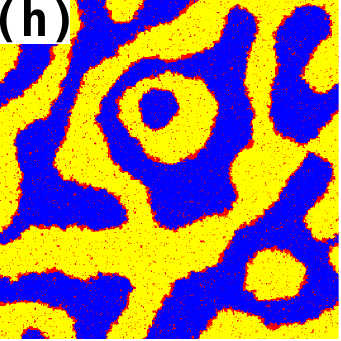}
\\[0.3cm]
\includegraphics[width = 0.09\textwidth]{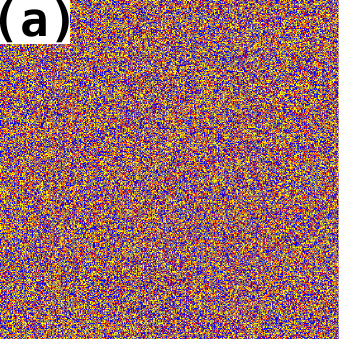} &
\includegraphics[width = 0.09\textwidth]{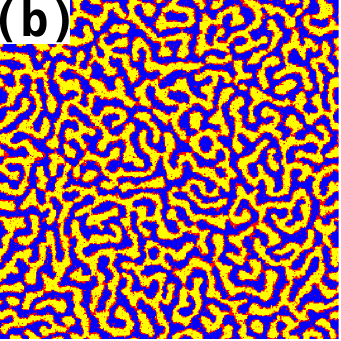} &
\includegraphics[width = 0.09\textwidth]{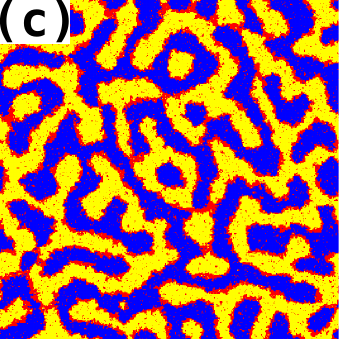} &
\includegraphics[width = 0.09\textwidth]{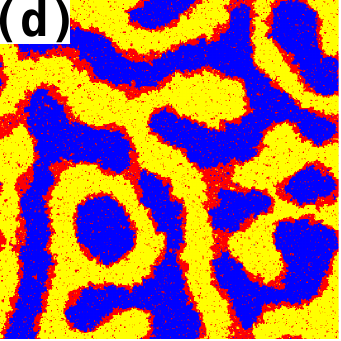}
\\[0.3cm]
\includegraphics[width = 0.09\textwidth]{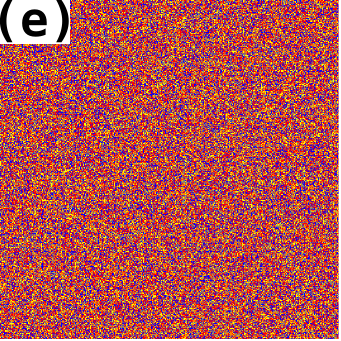} &
\includegraphics[width = 0.09\textwidth]{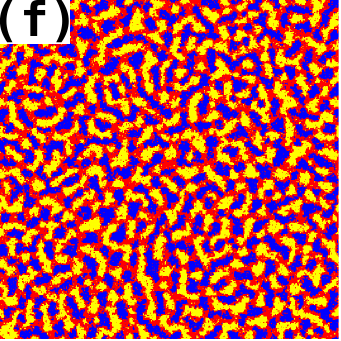} &
\includegraphics[width = 0.09\textwidth]{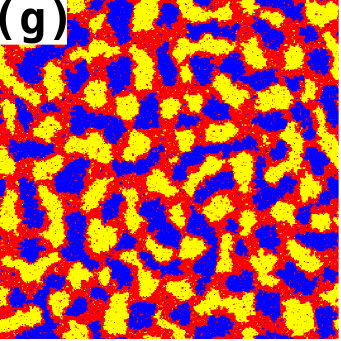} &
\includegraphics[width = 0.09\textwidth]{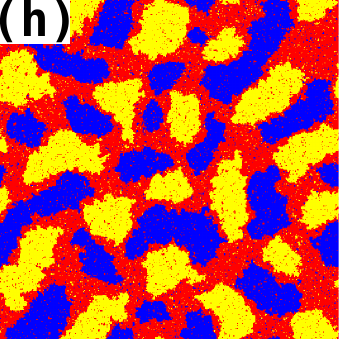}
\\[0.3cm]
\includegraphics[width = 0.09\textwidth]{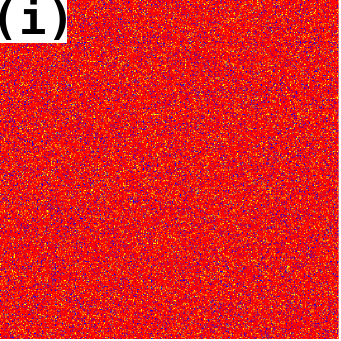} &
\includegraphics[width = 0.09\textwidth]{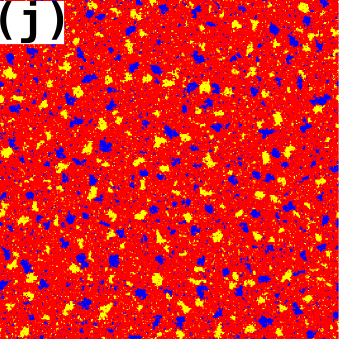} &
\includegraphics[width = 0.09\textwidth]{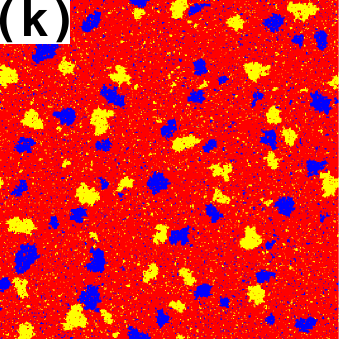} &
\includegraphics[width = 0.09\textwidth]{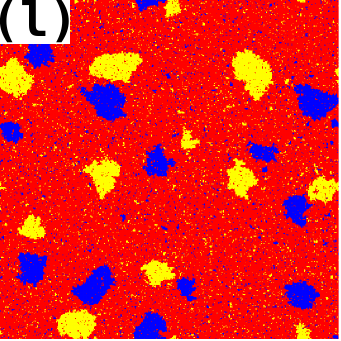}
\end{tabular}
\caption{Morphology formation and domain coarsening of the three
state model without evaporation for $L=512$ and $\beta = 1.0$.
From the top to the bottom $c_0=0.0,0.1,0.2,0.4,0.8$.
From the left to the right $t=0,8102,98714,729415$ MCS.}
\label{fig:PottsMorphology}
%\label{fig:PottsMorphologySlow}
\end{figure}
%\begin{figure}[!ht]
%\centering
%\begin{tabular}{cccc}
%\includegraphics[width = 0.09\textwidth]{fig-mkcsmc04a-stamped.pdf} &
%\includegraphics[width = 0.09\textwidth]{fig-mkcsmc04b-stamped.pdf} &
%\includegraphics[width = 0.09\textwidth]{fig-mkcsmc04c-stamped.pdf} &
%\includegraphics[width = 0.09\textwidth]{fig-mkcsmc04d-stamped.pdf}
%\\[0.3cm]
%\includegraphics[width = 0.09\textwidth]{fig-mkcsmc04e-stamped.pdf} &
%\includegraphics[width = 0.09\textwidth]{fig-mkcsmc04f-stamped.pdf} &
%\includegraphics[width = 0.09\textwidth]{fig-mkcsmc04g-stamped.pdf} &
%\includegraphics[width = 0.09\textwidth]{fig-mkcsmc04h-stamped.pdf}
%\\[0.3cm]
%\includegraphics[width = 0.09\textwidth]{fig-mkcsmc04i-stamped.pdf} &
%\includegraphics[width = 0.09\textwidth]{fig-mkcsmc04j-stamped.pdf} &
%\includegraphics[width = 0.09\textwidth]{fig-mkcsmc04k-stamped.pdf} &
%\includegraphics[width = 0.09\textwidth]{fig-mkcsmc04l-stamped.pdf}
%\end{tabular}
%\caption{As in Figure~\ref{fig:PottsMorphologySlow}
%for
%$c_0=0.2,0.4,0.8$ (from the top to the bottom).}
%\label{fig:PottsMorphology}
%\end{figure}

The two panels show that in the two state case, namely, for $c_0=0$, growth is very slow and in the time interval considered in the simulation, domains have just started to form.
\AM{This simply means that the scaling regime has not yet been reached, and that the overall process is still in its incipient phase.}
This line of reasoning is supported by comparing the configurations shown in the upper two rows of Fig.~\ref{fig:PottsMorphology}.
Indeed, looking at the first row, configurations referring to the case $c_0=0.0$ are shown.
\AM{They point out clearly} that the growth regime has not yet started, compared to the second row with $c_0 = 0.1$, where the domains have grown substantially.

Both \AM{analysis} techniques \AM{agree when predicting} that the domain growth is much faster when a moderate amount of zeroes is present in the system. This \AM{situation} is not unprecedented in literature, see, e.g., \cite{vacancies-paper} where the authors demonstrate a process of greatly speeding up the Ising Kawasaki dynamics by introducing one or several vacancies in the lattice. \AM{We argue} that the zero component of the three state model acts as a form of vacancy, since its interface cost is smaller.
\AM{Hence},  replacing it by a minus or \AM{by} a plus spin is cheaper.
Thus, zero sites are more mobile and greatly speed up the dynamics.
Indeed, for $c_0=0.1$, which is reported in the second row of the same figure, the domain formation started at about $10^3$ MCS.
This is also very well confirmed by the data in the right panel of Fig.~\ref{fig:Potts}: the curve referring to the case $c_0=0.1$ mildly grows \AM{until $10^3$ MCS} where it experiences an abrupt change \AM{of the slope} to a growth regime with the exponent $1/3$.

Both the correlation function and the structure factor measure characteristic domain sizes. \AM{Interestingly,} they give different results for $c_0 \geq 0.2$; see again Fig.~\ref{fig:Potts}.
The correlation function\AM{--based} measure is compatible with a $1/3$ exponent in the scaling regime for all values of $c_0$.
On the other hand, the structure factor\AM{--based} measure suggests that growth is slower when the zeroes concentration is increased.
The configurations plotted in the lower three rows of Fig.~\ref{fig:PottsMorphology} \AM{indicate} that the \AM{growth mechanisms} change when more zeroes are present.
While in the second row of Fig.~\ref{fig:PottsMorphology} zeroes form a thin film around plus and minus domains, so that growth happens essentially as in \AM{a two--state lattice} system, in the plots provided in the lower three rows of Fig.~\ref{fig:PottsMorphology} the situation is rather different:  domains of zeroes have sizes comparable with pluses and minus ones. \AM{Furthermore,} the three species \AM{seem to} compete during growth.
In the last row, when the zero concentration is very high, the process seems to \AM{become} more peculiar, in the sense that plus and minus domains grow inside a connected background of zeroes.
It is worth mentioning that similar behaviors have been \AM{observed} in different regimes, specifically in the study of the \AM{metastability occurring in the framework of}  the Blume--Capel model \cite{CN2013,CNS2017,CO1996,LL2016}, when growth does not happen via \AM{the} coalescence of small droplets \AM{but rather via a} sudden nucleation of a large droplet.

The fact that different values of the zeroes concentration give rise to different growing mechanisms suggests that, for this three state system, the way in which the structure factor measures the size of the domains is more reliable than that provided by the correlation function technique. Even if, from the experimental point of view, this case might seem less interesting, our results can be related to situations where the concentration of the active components in the solvent does not vary much, \AM{as it would be for instance the case} for a certain short amount of time at the bottom of the film or for \AM{a} very slow evaporation.

\subsection{Domain growth in the three state model with evaporation}

\begin{figure}
\centering
\begin{tabular}{cc}
\includegraphics[width = 0.22\textwidth]{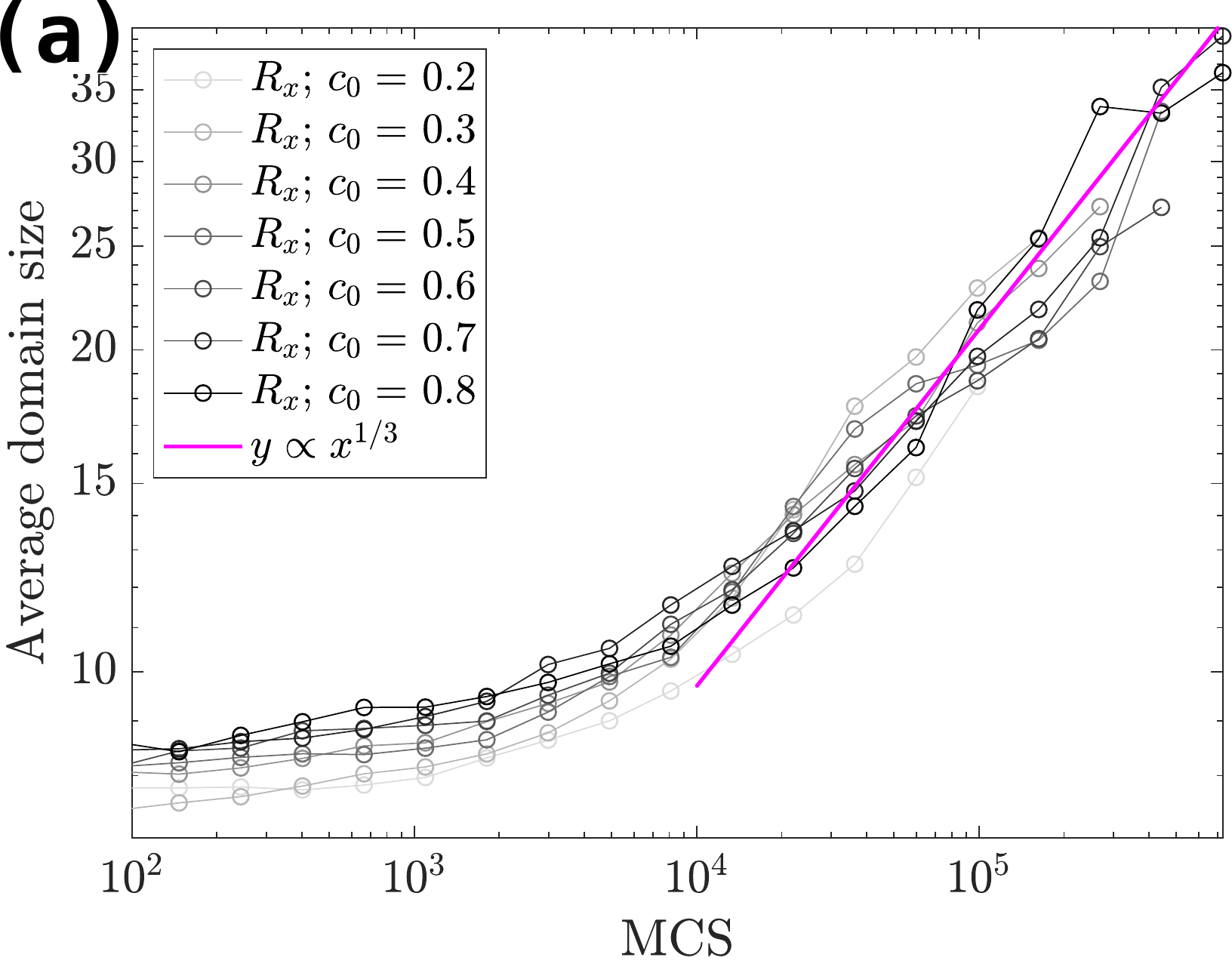}
&
\includegraphics[width = 0.218\textwidth]{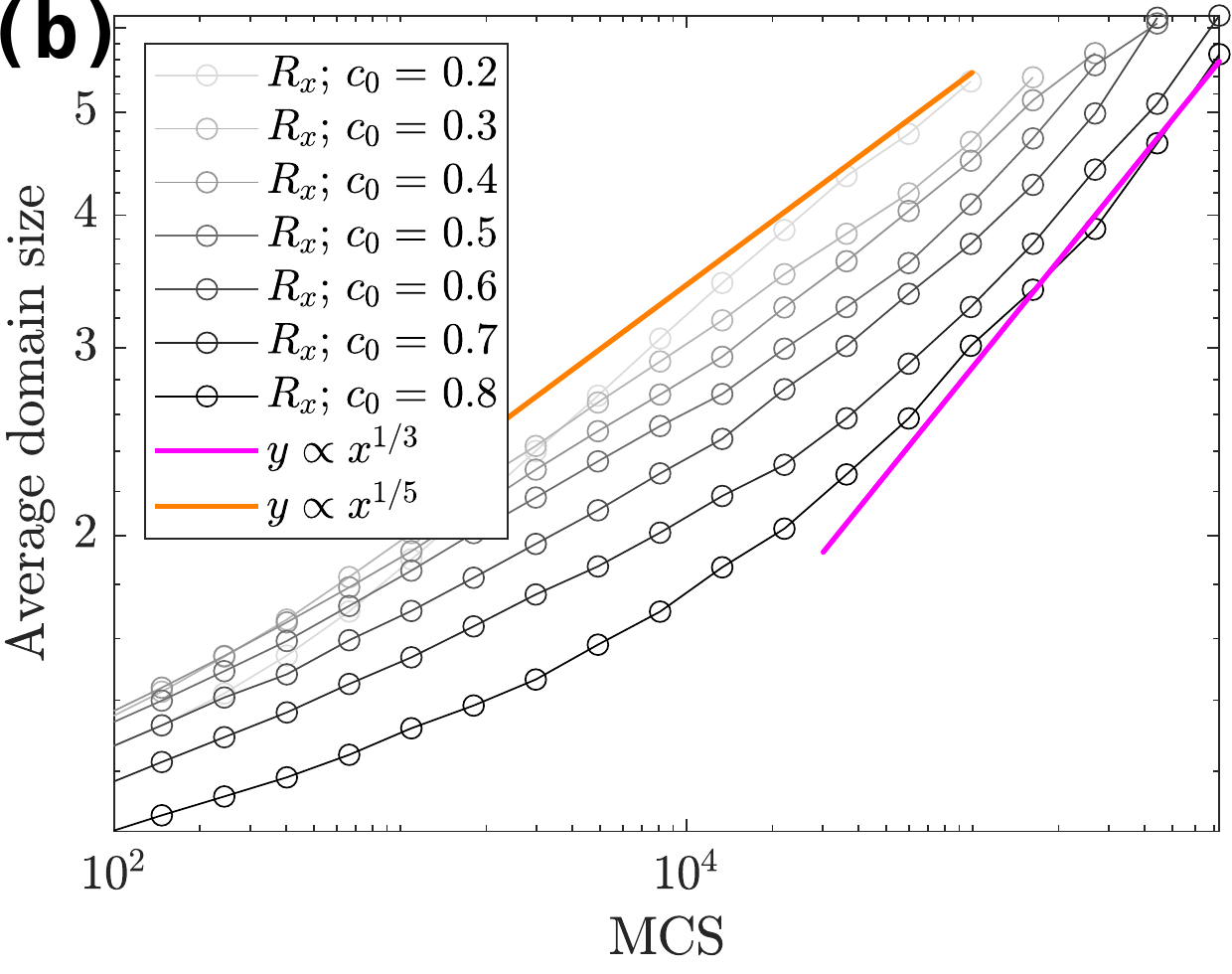}
\end{tabular}
\caption{Horizontal domain size--based measure via the correlation function (on the left) and the structure function computation (on the right) for the three--state lattice model with evaporation for $L = 512$ and $\beta = 1.0$. One single system history per initial concentration.}
\label{fig:evapH}
\end{figure}

\begin{figure}
\centering
\begin{tabular}{cc}
\includegraphics[width = 0.22\textwidth]{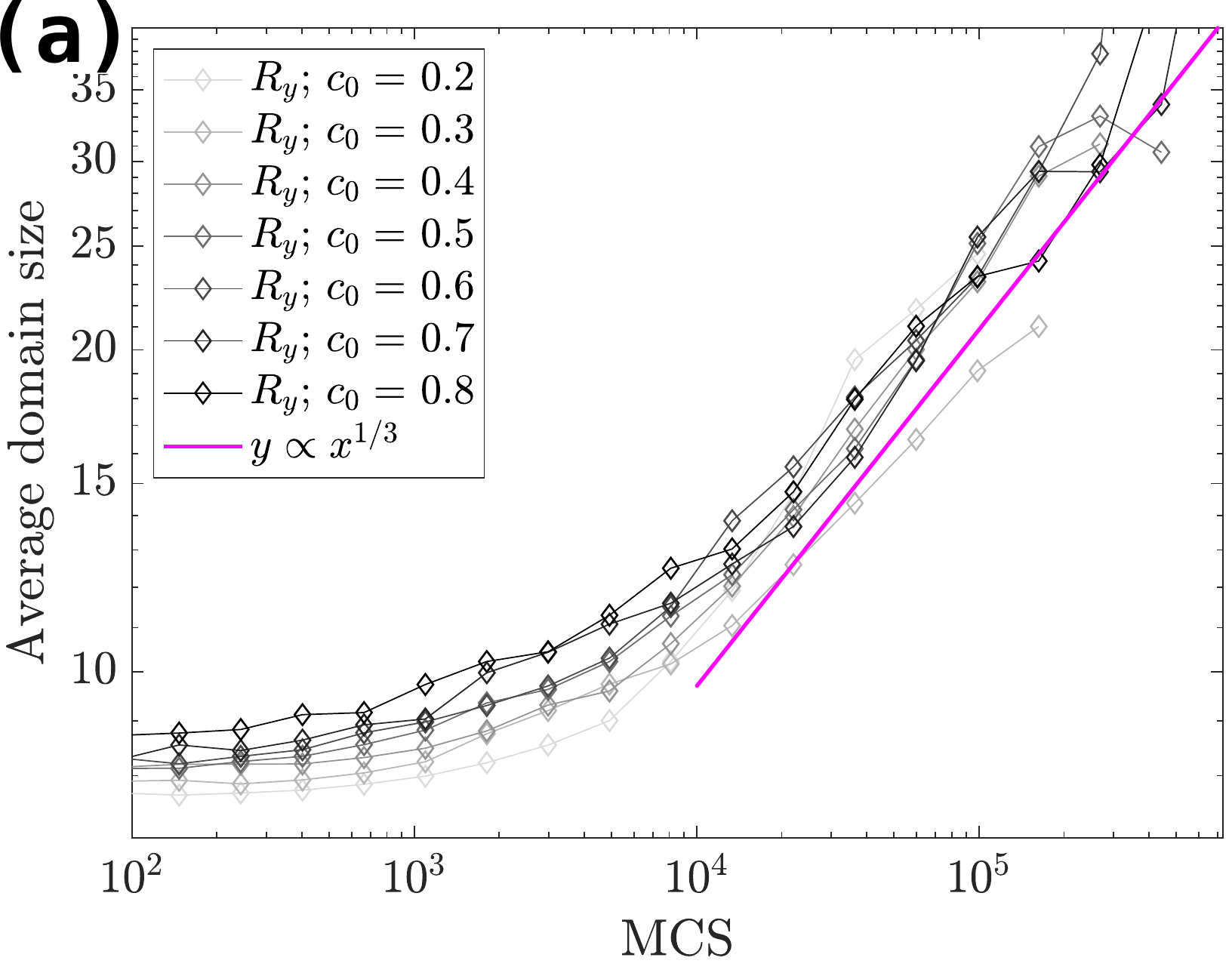}
&
\includegraphics[width = 0.218\textwidth]{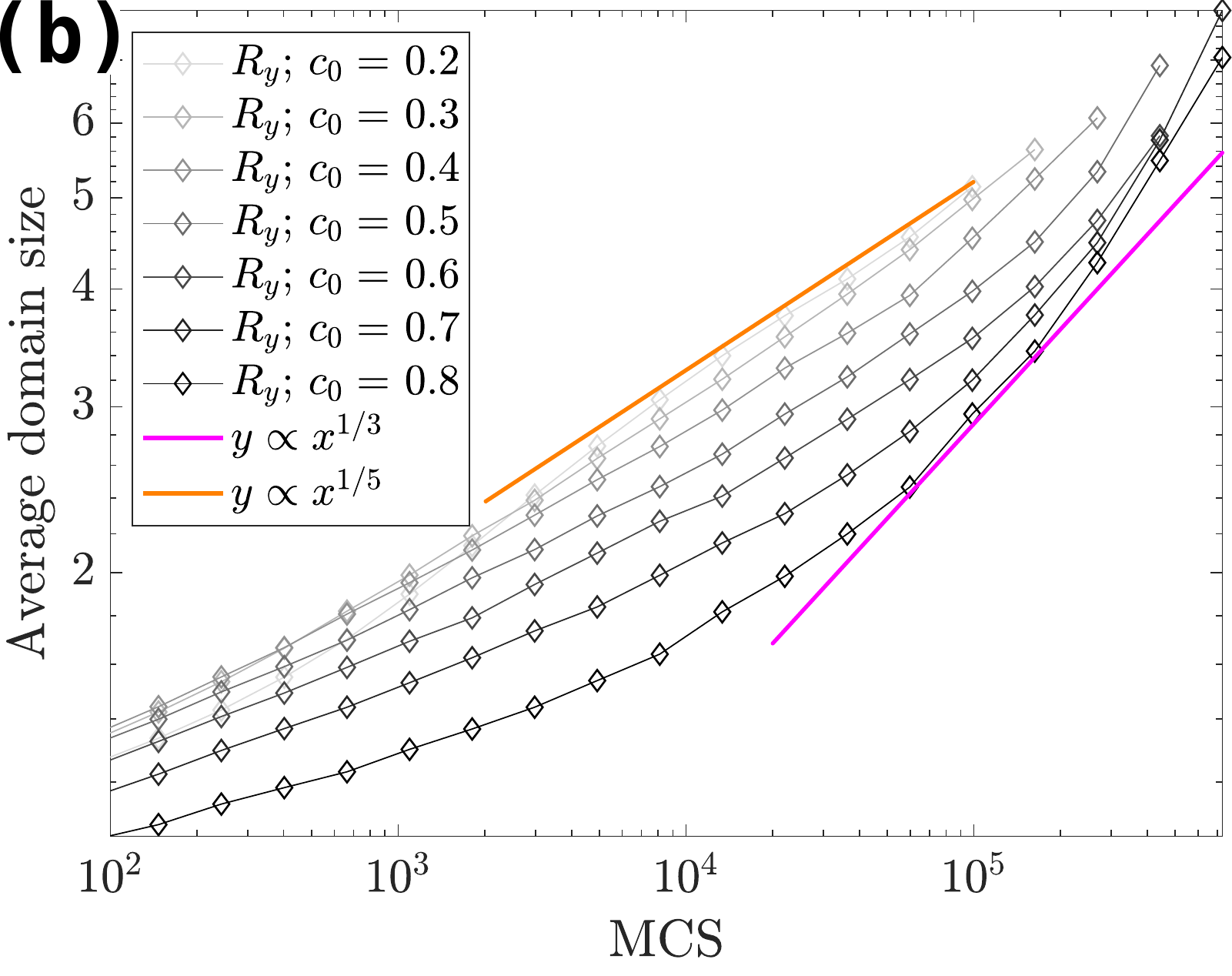}
\end{tabular}
\caption{As in Fig.~\ref{fig:evapH} for the vertical domain size.}
\label{fig:evapV}
\end{figure}

Now, we discuss our results for the domain growth in the presence of the evaporation, which is the most interesting case \AM{in terms of applications to ternary mixtures in the context of organic solar cells referred to in the introduction}.
The horizontal and vertical domain sizes measured with the correlation function and with the structure factor are reported in Fig.~\ref{fig:evapH} and Fig.~\ref{fig:evapV}, respectively.
Just like in the case discussed in the previous section, the two--point correlation function shows a domain growth exponent of $1/3$ in the appropriate regime, while the structure factor method gives a more complex behaviour, showing growth exponents between $1/5$ and $1/3$ for $R_x$.
\AM{What concerns} $R_y$, we see briefly a growth exponent of $1/3$, which then speeds up further near the end of the evaporation.
This asymmetry is not necessarily surprising as this problem is anisotropic due to the evaporation of the zeroes at the top--row of the lattice.

\begin{figure}[!ht]
\centering
\begin{tabular}{cccc}
\includegraphics[width = 0.09\textwidth]{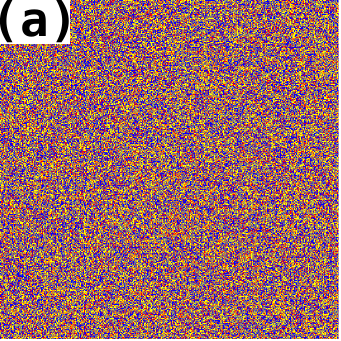} &
\includegraphics[width = 0.09\textwidth]{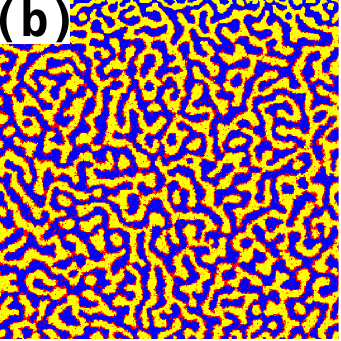} &
\includegraphics[width = 0.09\textwidth]{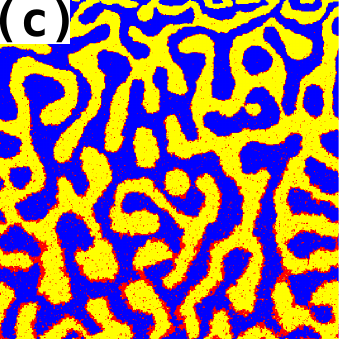} &
\\[0.3cm]
\includegraphics[width = 0.09\textwidth]{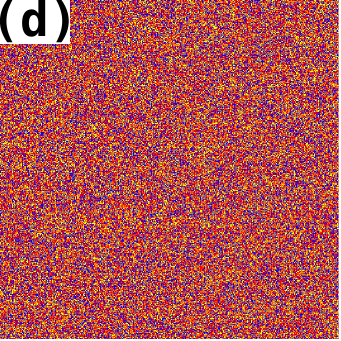} &
\includegraphics[width = 0.09\textwidth]{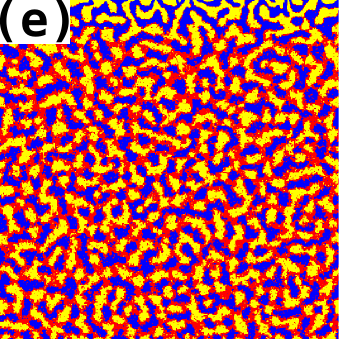} &
\includegraphics[width = 0.09\textwidth]{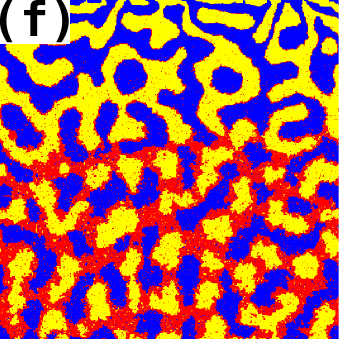} &
\\[0.3cm]
\includegraphics[width = 0.09\textwidth]{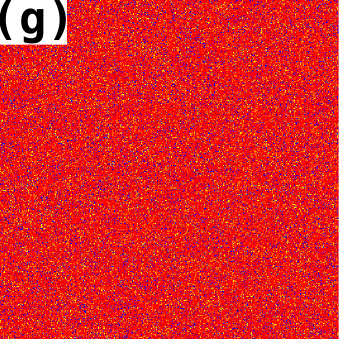} &
\includegraphics[width = 0.09\textwidth]{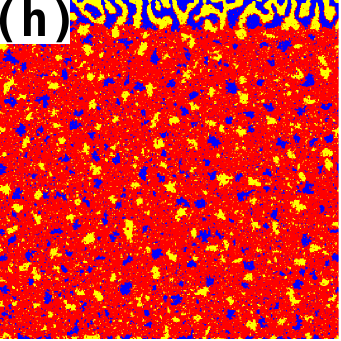} &
\includegraphics[width = 0.09\textwidth]{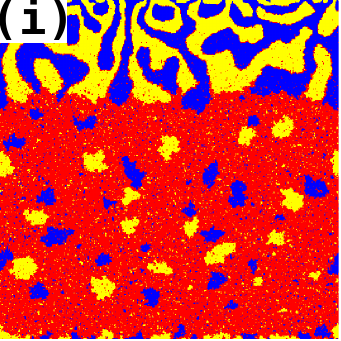} &
\includegraphics[width = 0.09\textwidth]{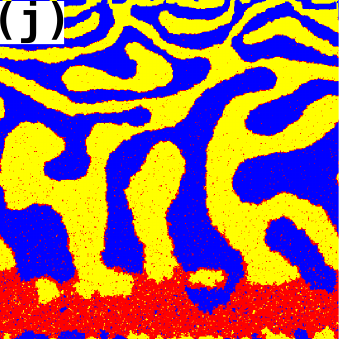}
\end{tabular}
\caption{Morphology formation and domain coarsening of the three--state lattice model with evaporation for $L=512$ and $\beta = 1.0$.
From the top to the bottom $c_0=0.2,0.4,0.8$.
From the left to the right $t=0,8102,98714,729415$ MCS.}
\label{fig:evapMorphology}
\end{figure}

Even in the presence of evaporation, the growth mechanism seems to depend on the solvent concentration, see Fig.~\ref{fig:evapMorphology}. Note that the last frame (for  $729415$ MCS) is missing for the first two initial concentrations of zeroes, as the length of the simulation is decided by the final concentration of zeroes (recall the stopping condition for the dynamics is when $c_0 = 0.1$) and, in these cases,  that value is reached for shorter simulation lengths. By similar arguments as before, it is believed that the structure factor yields the most meaningful domain size calculation, which is further supported by the observation that $R_y > R_x$ near the end of the evaporative process (compare Fig.~\ref{fig:evapH} and Fig.~\ref{fig:evapV}).
This also appears to be the case when looking at the final configurations in Fig.~\ref{fig:evapMorphology}, and it is thus believed to be a good indicator of the validity of the domain size calculations.

\subsection{Effect of temperature}
\AM{Temperature is an important factor in the context of our lattice models as well as for the actual experimental processing of the thin film, both in the initial phase of the solvent evaporation, but also in the late stages and even for after processing, via thermal annealing.
Direct comparisons of the simulation results with experiment are not feasible at this stage due to lack of temperature controlled in-situ experimental data especially at early stages but also due to the qualitative character of the temperature in our model, captured only in terms of $\beta^{-1}$.
Nevertheless, we can see clear tendencies on the domains growth and this is a good starting point for further investigations.} In this section, we investigate the effect the temperature $\beta^{-1}$ has on the domains growth \AM{as they are formed in the context of} the three--state lattice model without and with evaporation. The estimates of the domains size are done via the structure factor method.

\begin{figure}[!ht]
\centering
\begin{tabular}{cc}
\includegraphics[width = 0.22\textwidth]{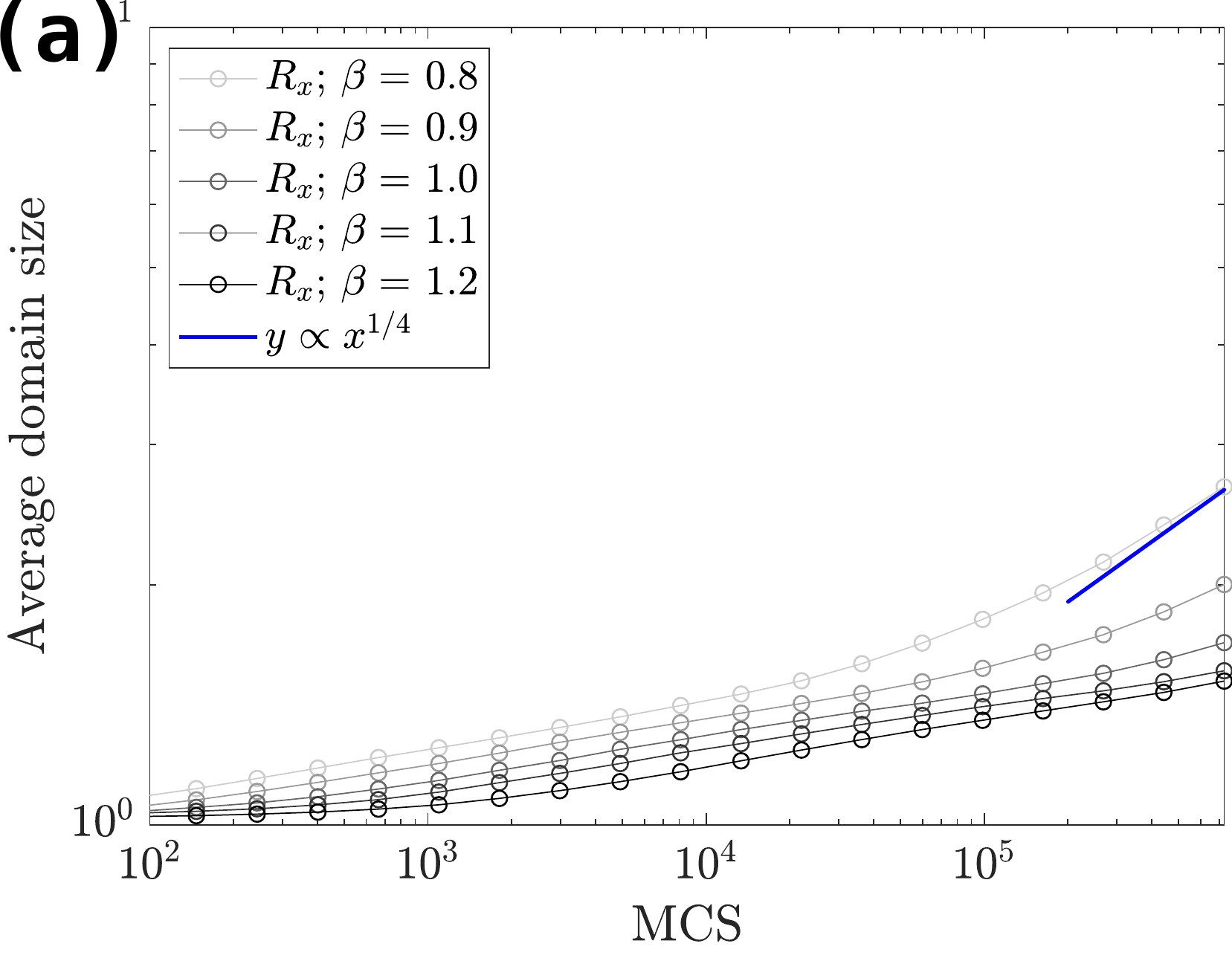} &
\includegraphics[width = 0.22\textwidth]{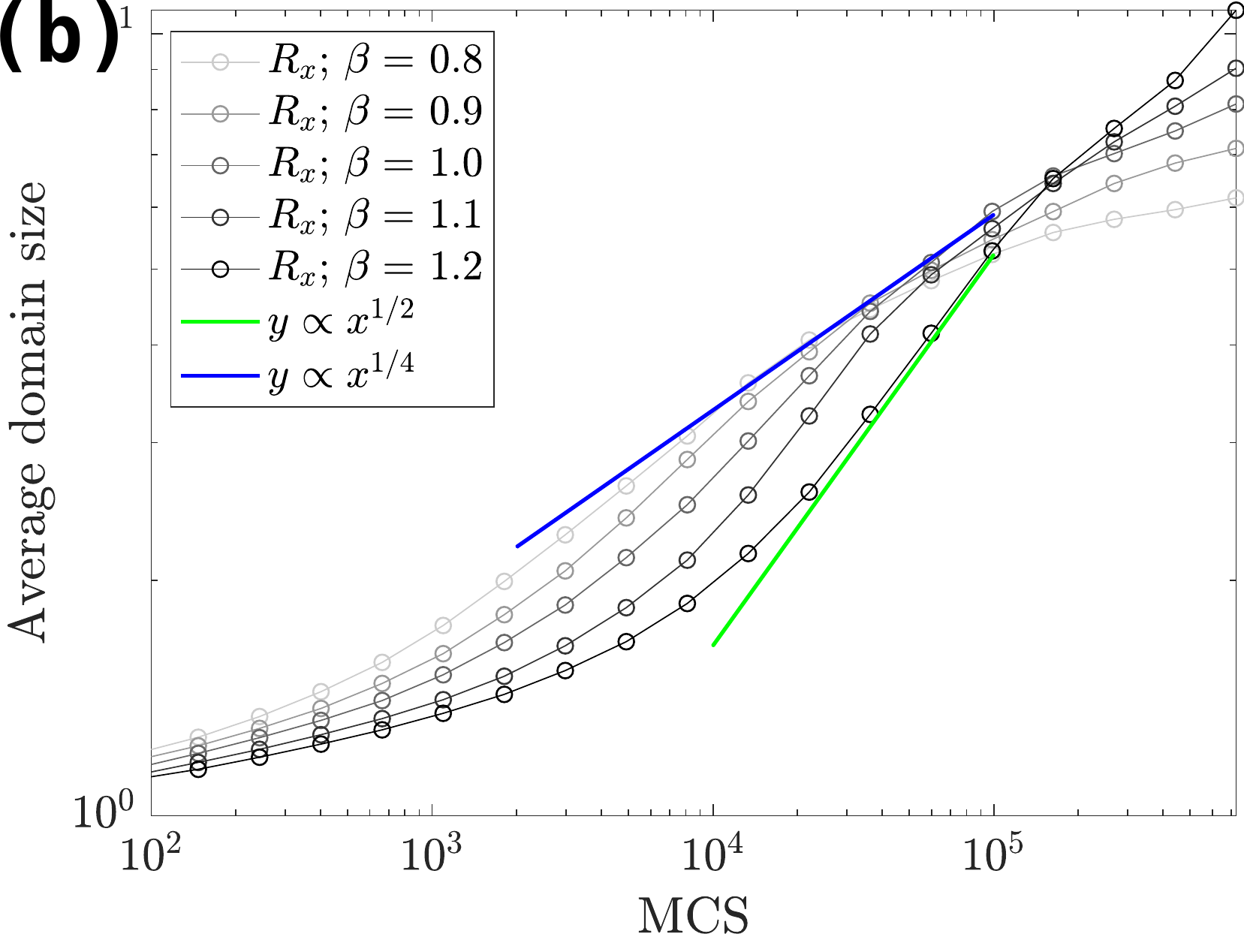}
\\[0.3cm]
\includegraphics[width = 0.22\textwidth]{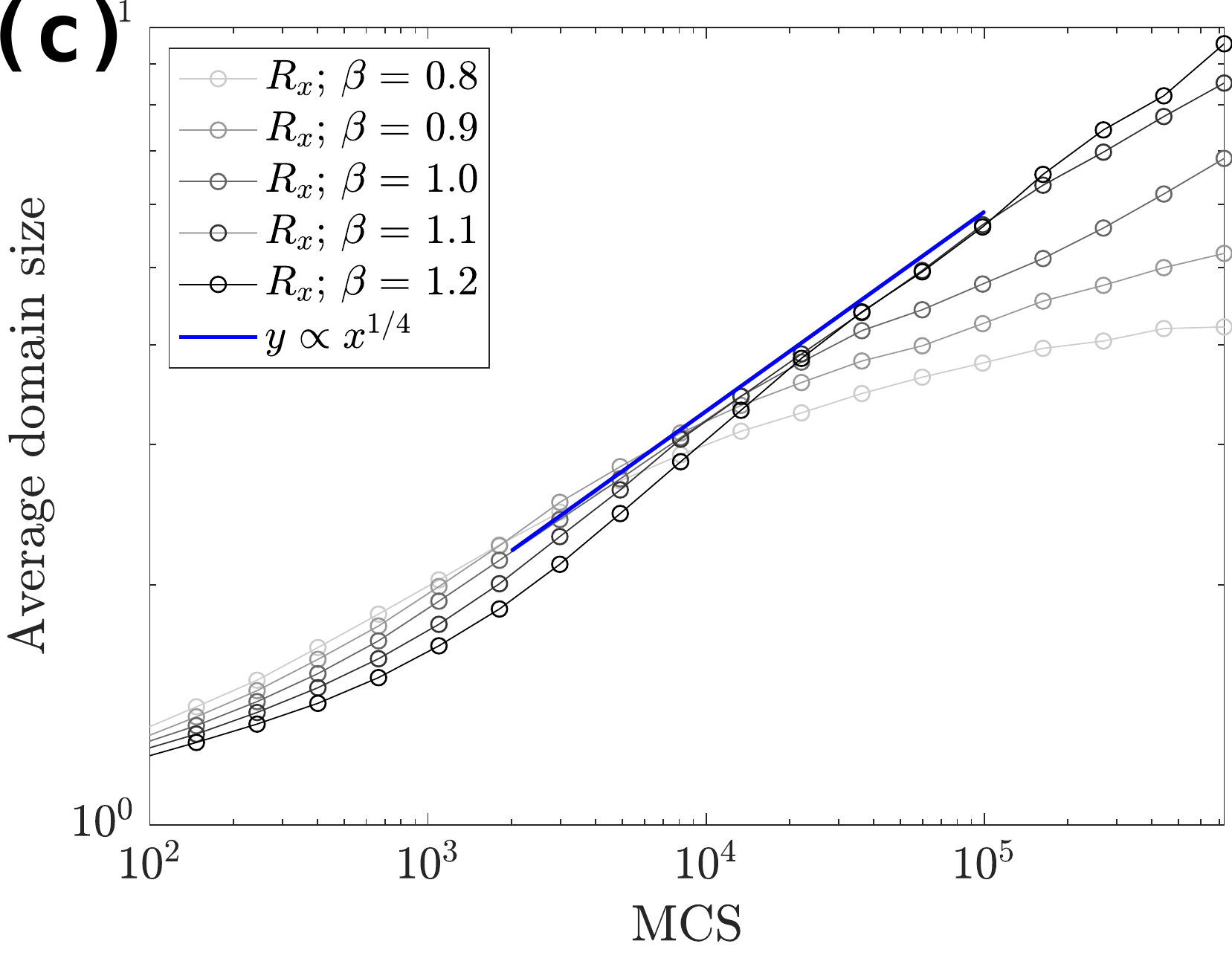} &
\includegraphics[width = 0.22\textwidth]{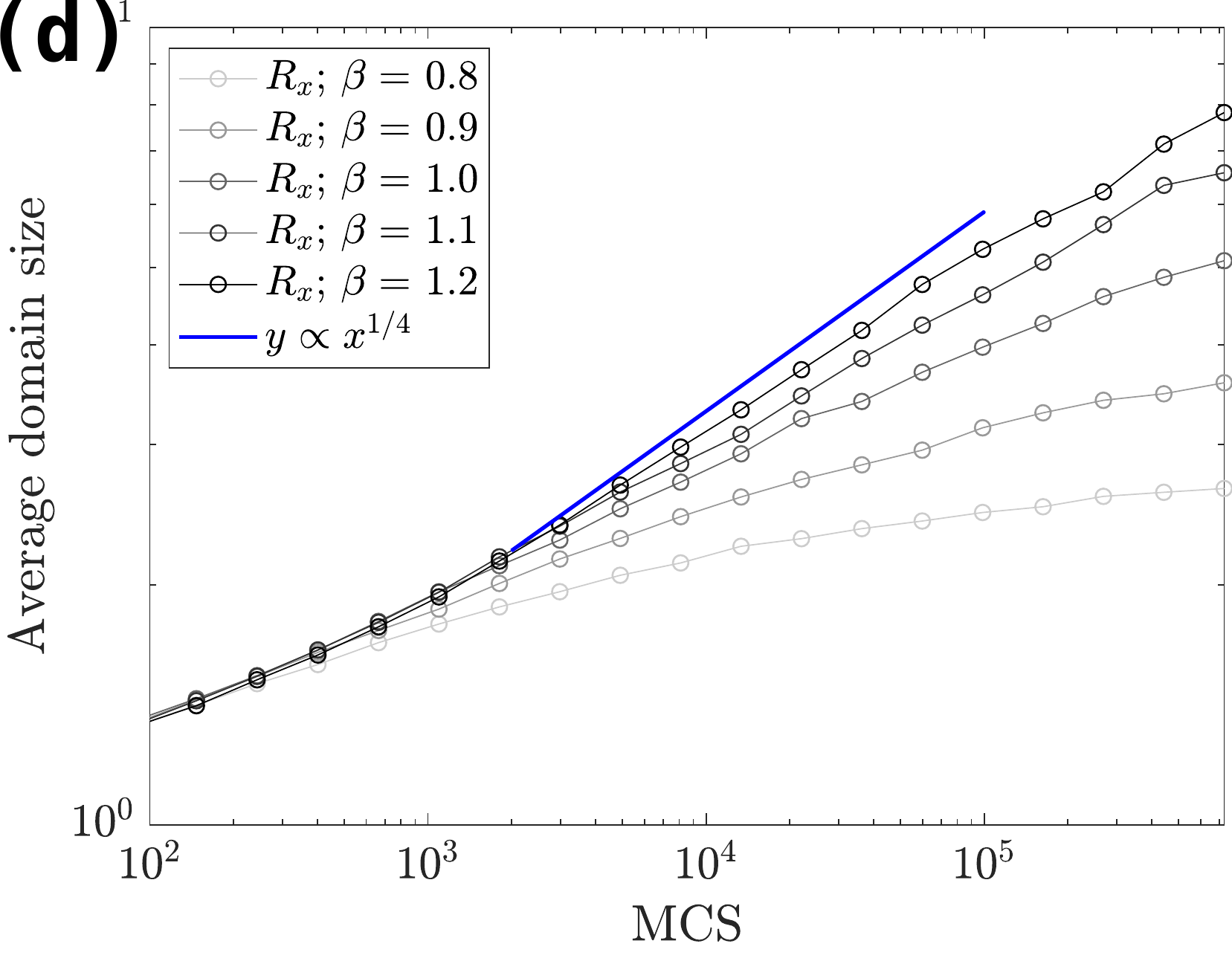}
\end{tabular}
\caption{Structure factor estimate of horizontal domain size
for the three--state lattice model without evaporation
for $c_0=0,0.2,0.4,0.6$ (lexicographical order)
for different values of the temperature (see inset).}
\label{fig:tempDep}
\end{figure}

In Fig.~\ref{fig:tempDep}, we plot our result in absence of evaporation. Several noteworthy aspects appear.
With a binary $\pm1$ mixture, namely, $c_0=0.0$ (see the left top panel in the figure), it is clear that increasing temperature is associated with accelerating dynamics. As the concentration of zeroes is increased, this behaviour becomes more complex.
More specifically, the initial trend is still comparable, i.e. increasing temperature can be associated with larger domains after the same time.
At larger times, this is no longer the case, see \AM{the time slice} $\sim 10^5$ MCS in the top right panel of Fig.~\ref{fig:tempDep}, where the $\beta = 1.2$ data crosses the $\beta = 0.8$ one.
As the solvent concentration continues to increase, the effect of the temperature \AM{diminishes} before the inflection point (which also occurs at earlier times, \AM{bringing evidence to} the idea that the third species speeds up the dynamics).
\AM{Finally,  for} $c_0=0.6$,  the inflection point is no longer visible in the current time domain; see the right bottom panel in Fig.~\ref{fig:tempDep}.

\begin{figure}[!ht]
\centering
\begin{tabular}{cccc}
\includegraphics[width = 0.09\textwidth]{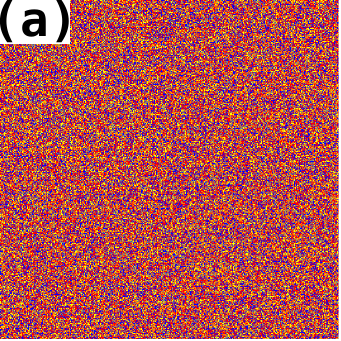} &
\includegraphics[width = 0.09\textwidth]{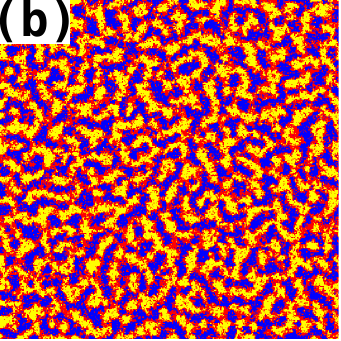} &
\includegraphics[width = 0.09\textwidth]{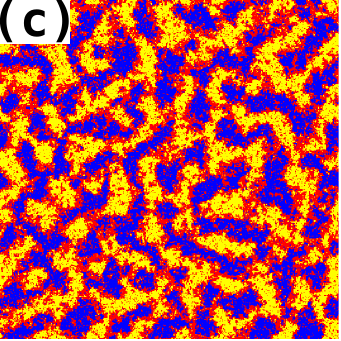} &
\includegraphics[width = 0.09\textwidth]{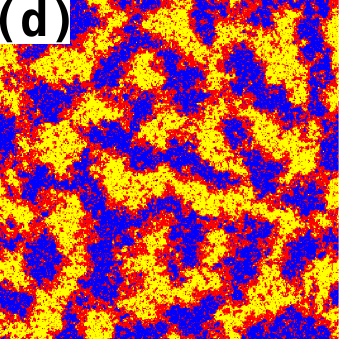}
\\[0.3cm]
\includegraphics[width = 0.09\textwidth]{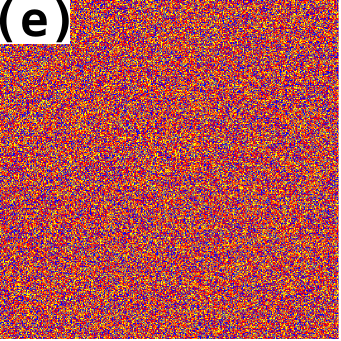} &
\includegraphics[width = 0.09\textwidth]{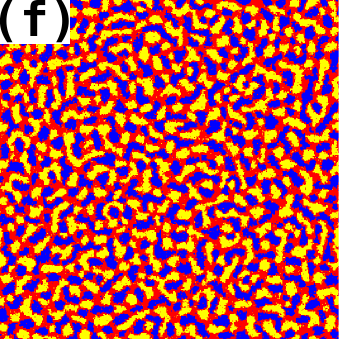} &
\includegraphics[width = 0.09\textwidth]{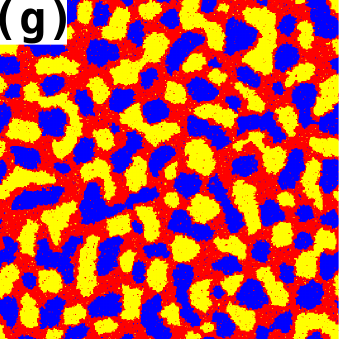} &
\includegraphics[width = 0.09\textwidth]{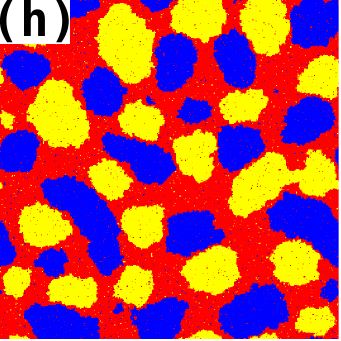}
\end{tabular}
\caption{Morphology formation and domain coarsening of the three--state lattice model without evaporation for $L=512$, $c_0=0.4$, 
$\beta = 0.8$ (top row), and 
$\beta=1.2$ (bottom row). 
From left to right: $t=0,8102,98714,729415$ MCS.}
\label{fig:tempDep-morphology}
\end{figure}

It is instructive to \AM{look for a moment at} the morphology formation exhibited in Fig.~\ref{fig:tempDep-morphology} \AM{for the case} $c_0 = 0.4$.
Note that \AM{for a} high temperature ($\beta = 0.8$), the zero sites (depicted in red) readily penetrate the minus-- and plus--filled domains (the yellow and blue regions), to the point where the interface boundaries are rather difficult to identify. \AM{When the temperature is low} (e.g. for $\beta = 1.2$), the situation is significantly different, and the phases are clearly defined, with minimal interpenetration.
It is difficult to say by inspection that the domains are larger in terms of area in the second case (i.e. lower row in Fig.~\ref{fig:tempDep-morphology}), but it is clear that they are more well defined.
Since the Fourier transform is quite sensitive to the sharpness of the boundaries of the domains, this will affect the results in Fig.~\ref{fig:tempDep} in the sense that results for high $\beta$ are more accurate compared to those for low $\beta$.

\begin{figure}[!ht]
\centering
\begin{tabular}{cc}
\includegraphics[width = 0.17\textwidth]{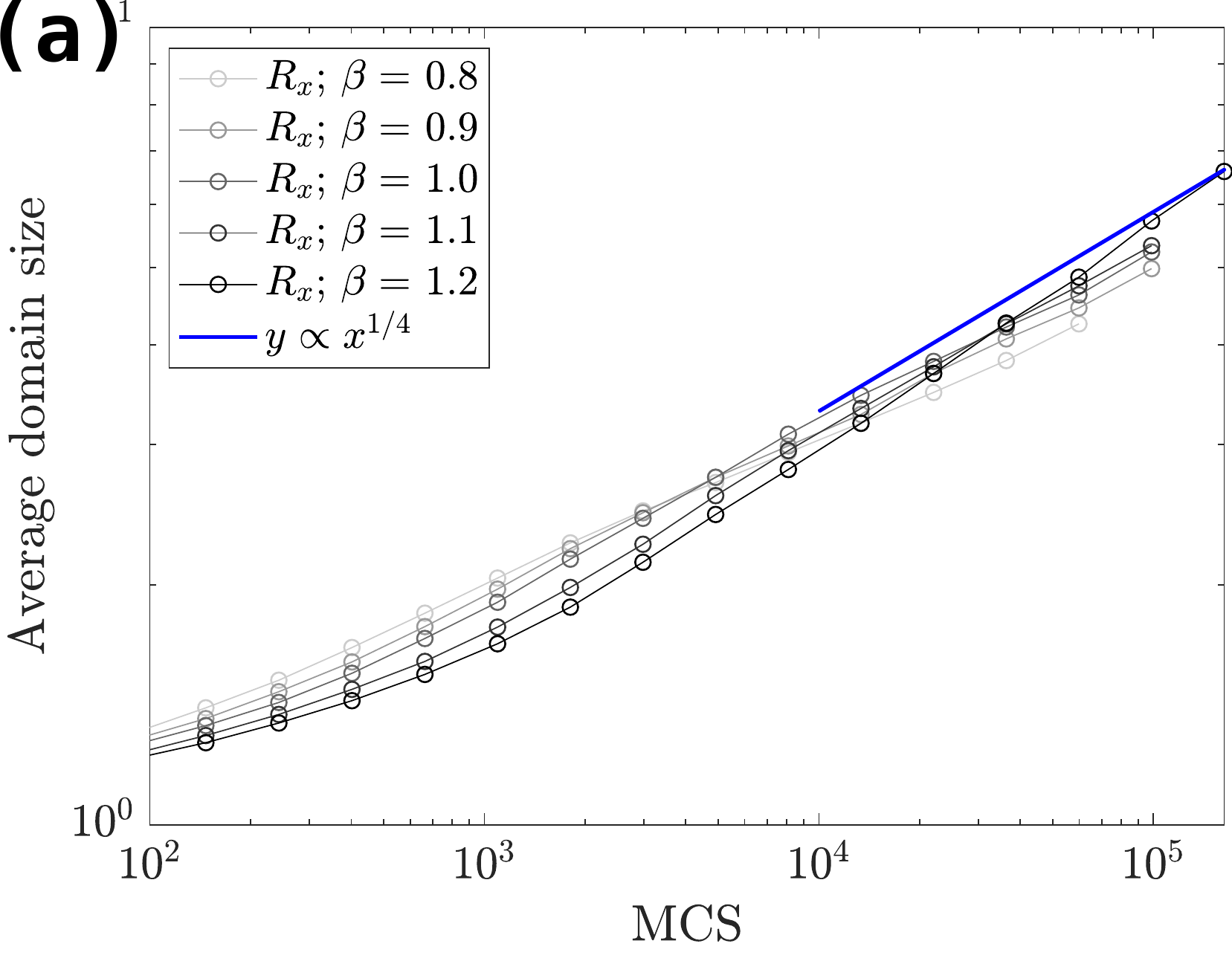} &
\includegraphics[width = 0.17\textwidth]{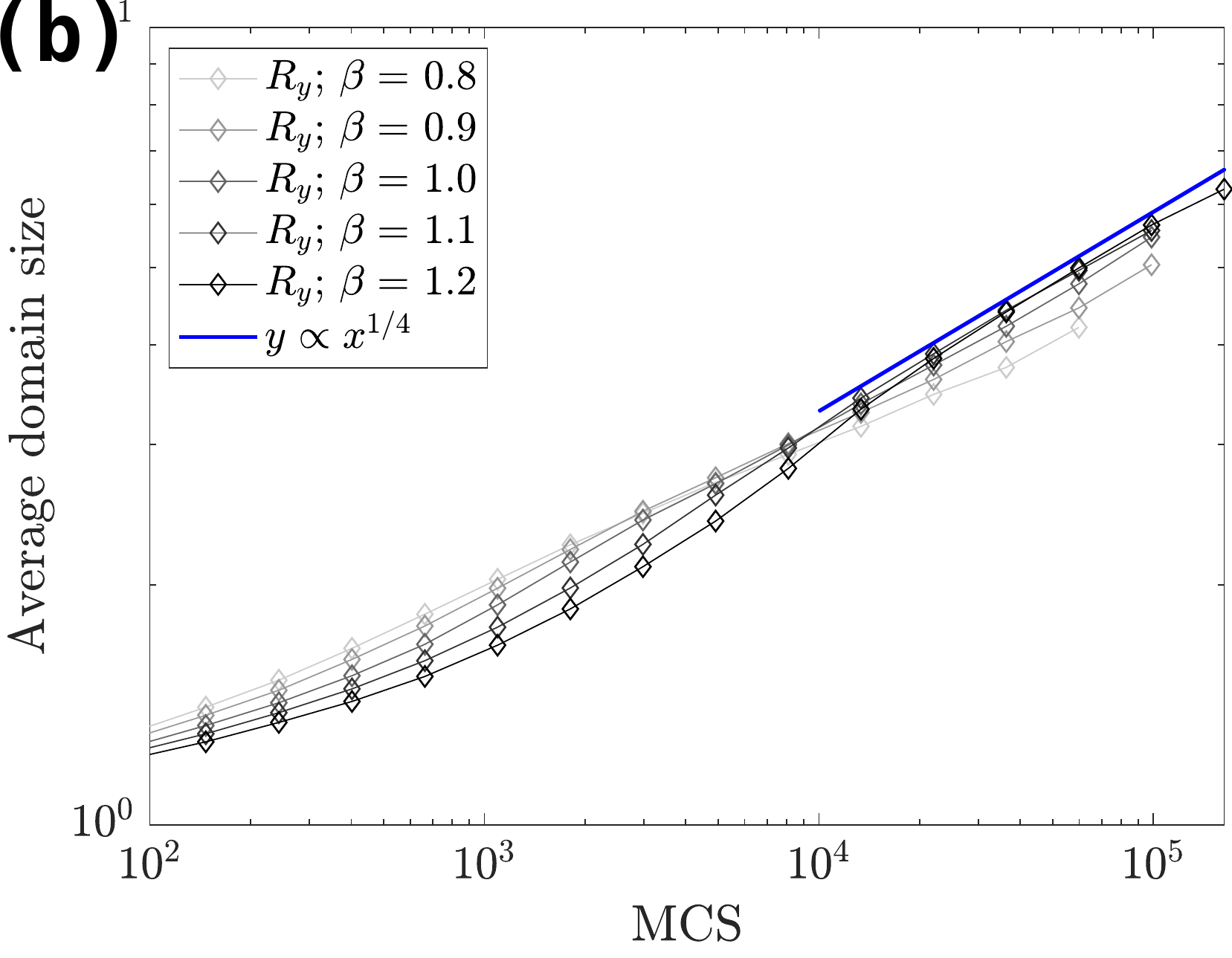}
\\[0.3cm]
\includegraphics[width = 0.17\textwidth]{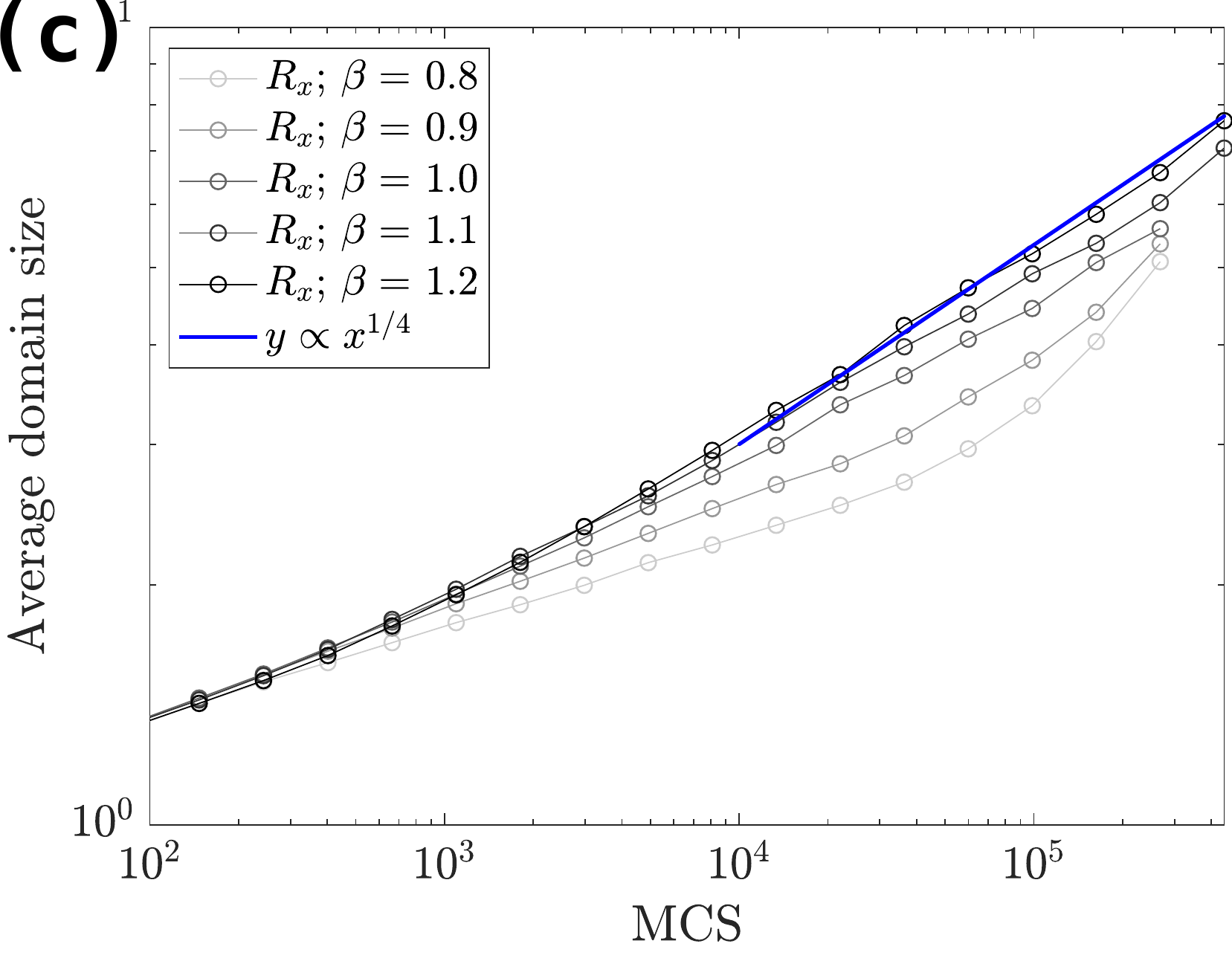} &
\includegraphics[width = 0.17\textwidth]{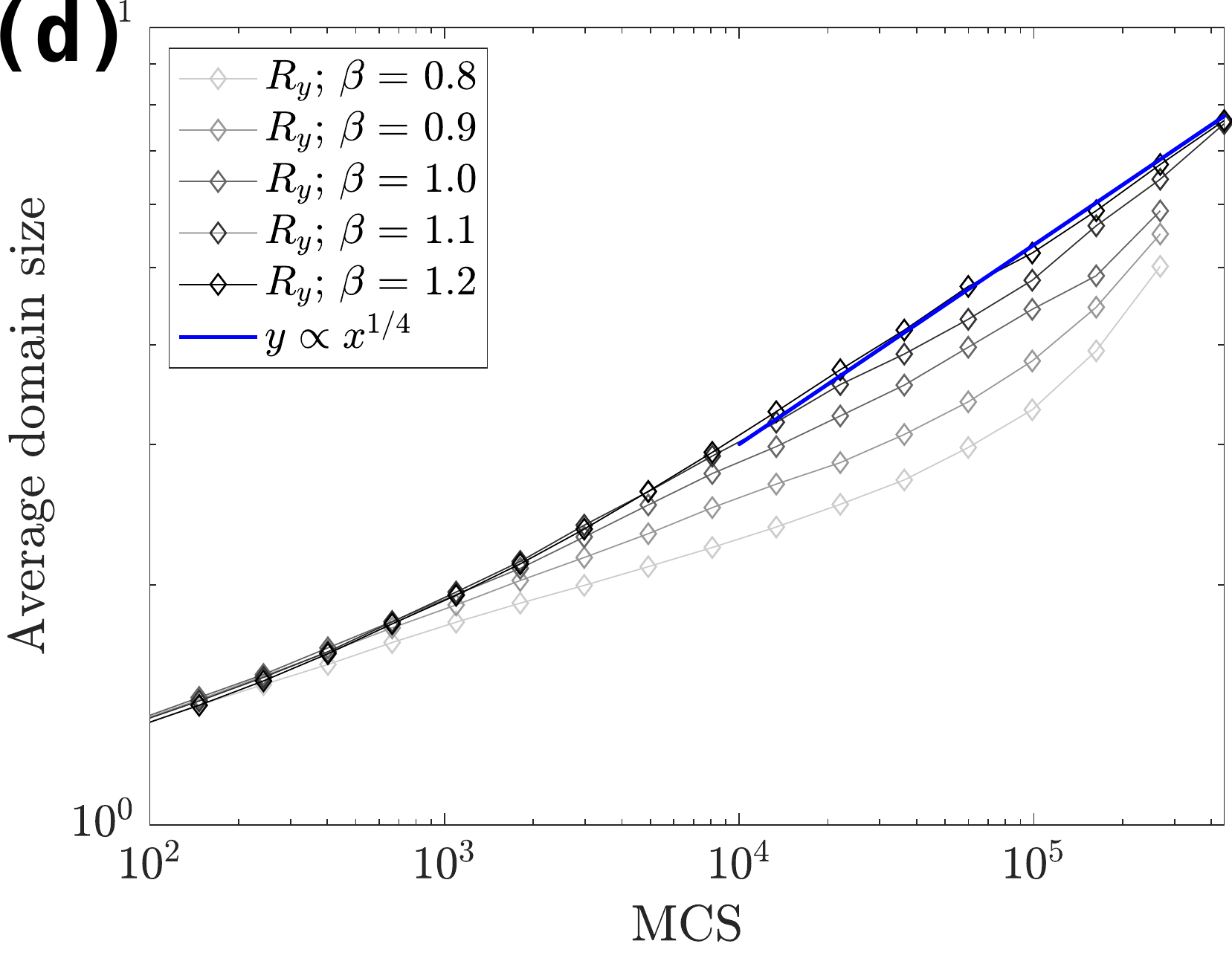}
\\[0.3cm]
\includegraphics[width = 0.17\textwidth]{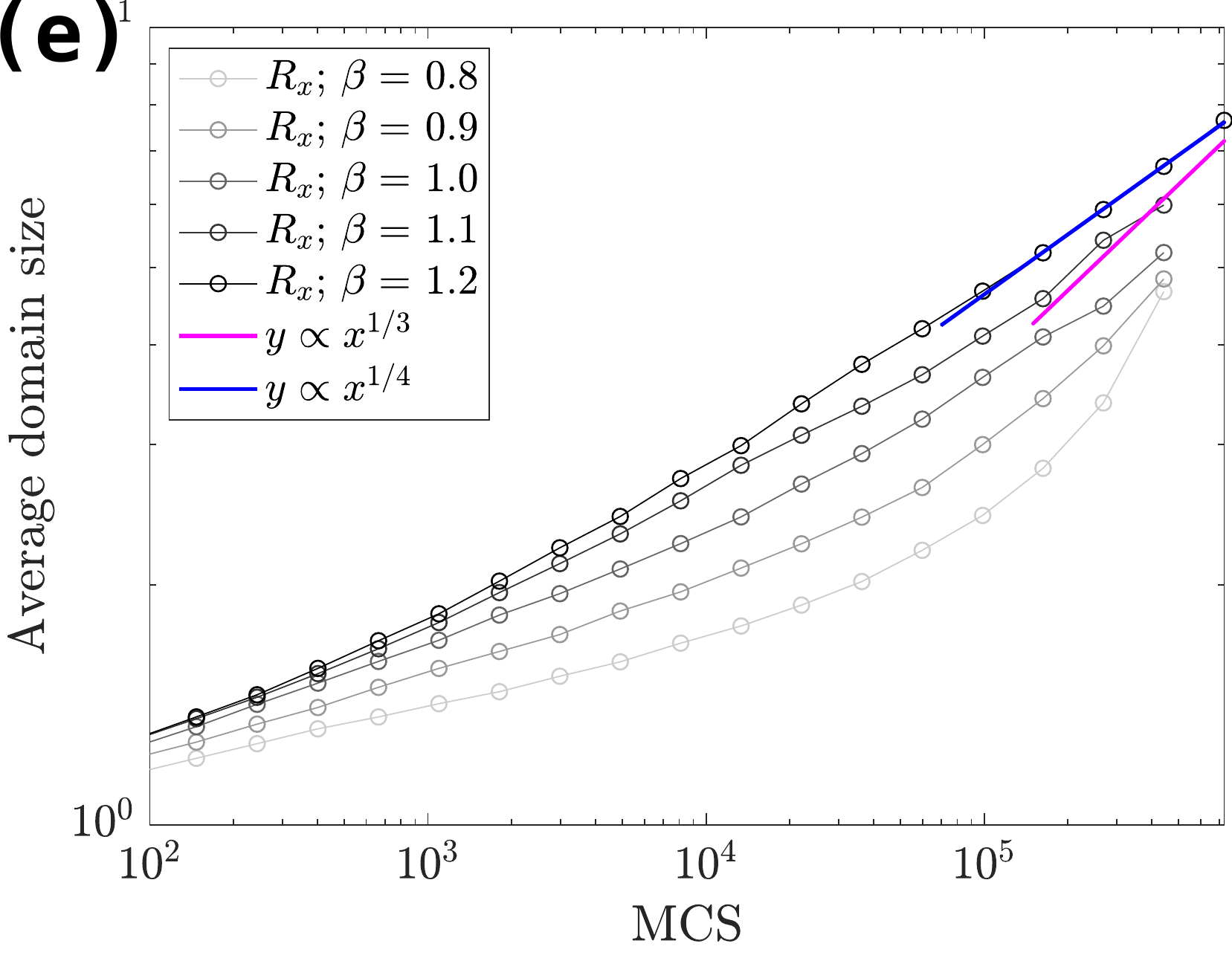} &
\includegraphics[width = 0.17\textwidth]{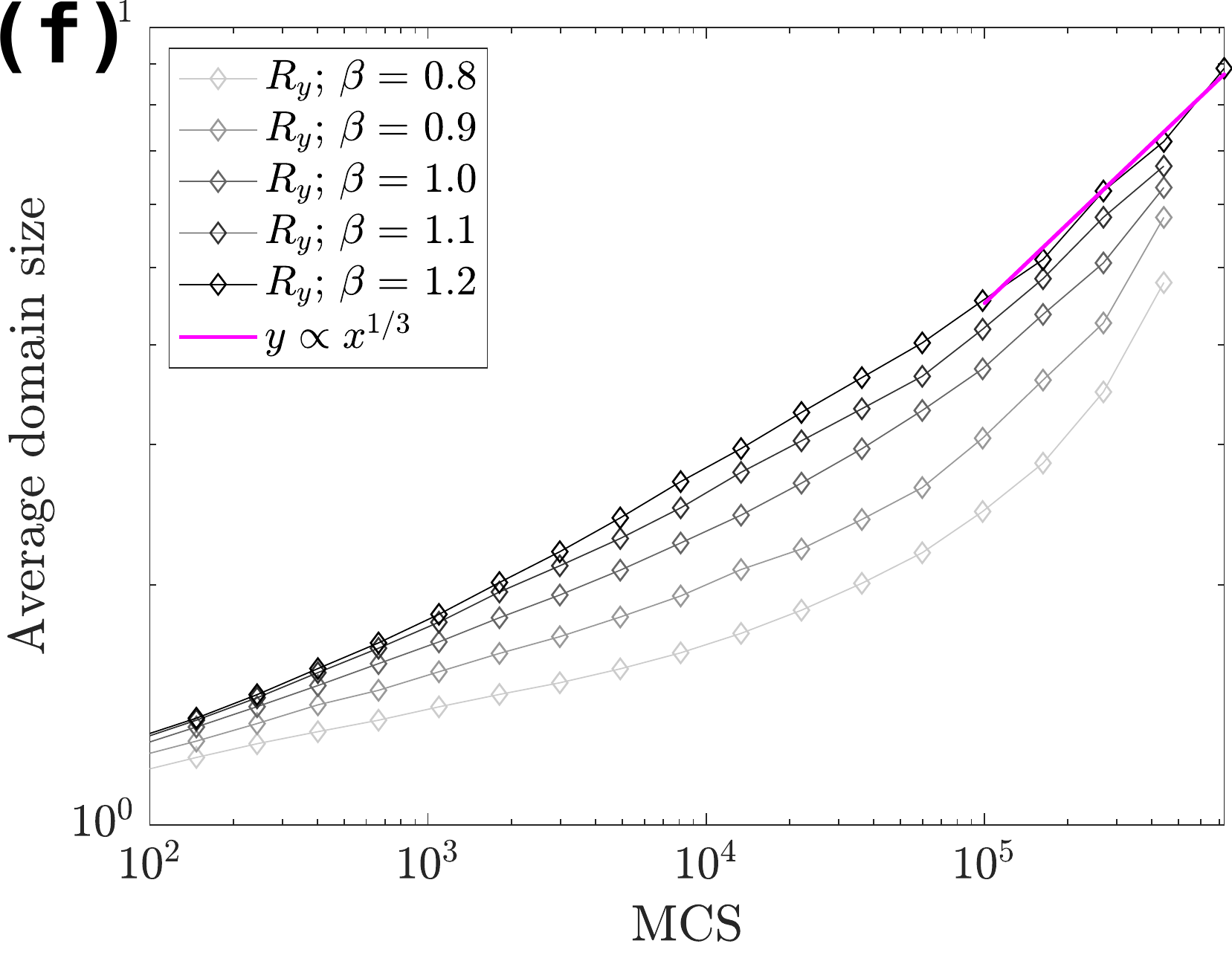}
\end{tabular}
\caption{Structure factor estimate of horizontal (left)
and vertical (right) domain size
for the three state model with evaporation
for $c_0=0.2,0.4,0.6$ (from top to bottom)
for different values of the temperature (see inset).}
\label{fig:tempDep-evap}
\end{figure}

We further consider the effect of temperature when the zeroes evaporate through the top--row of the lattice, see Fig.~\ref{fig:tempDep-evap}.
These data appear consistent with the case without evaporation \AM{at least from the point of view that} the initial behaviour is similar.
At longer times, the evaporation effects dominate, and the domain coarsening deviates from \AM{scenarios computed with} the three--state lattice model without evaporation.
Here again, we observe the inflection point, but only in the first row of Fig.~\ref{fig:tempDep-evap}, indicating that the dynamics are further accelerated due to the evaporation.

A proposed mechanism explaining the inflection points in Fig.~\ref{fig:tempDep} and Fig.~\ref{fig:tempDep-evap} is the following: high temperatures and/or high ratio of $\pm 1$ sites favours the growth of the domains. \AM{This seems to hold} until a certain critical domain size is reached, after which both the increase in temperature, and hence,  increase in fluctuations allow the zeros to penetrate the domains more easily. \AM{Consequently, this leads to an altering} of the average domain size.

\section{Concluding remarks}

	\AM{In this work, we proposed} a quantitative analysis of phase formation and domain growth in ternary mixtures, both with conservative and non--conservative dynamics. In the latter case, one component is evaporated from the top of the lattice in a process akin to that utilised in the fabrication of solution borne thin films used in photovoltaics \AM{applications based on organic solar cells}. We reproduce the 1/3 exponent for the Ising model, while for the
ternary--state model without evaporation and for that with evaporation we obtain lower values of this exponent. \AM{Estimating how domains  grow  is a quite complex task -- in our context, it is heavily influenced by the initial mixture concentration as well as by the temperature.} Interestingly, note that the morphologies obtained in the present simulations can be found as well in experimentally--built thin films; see e.g. \cite{Michels,Sprenger}.

\section{Further research}

	Suggestions for further research include an extension to three dimensions for improved applicability to experiments and to validate the two--dimensional results by studying a slice of the morphology.
	Staying in two dimensions, once could conceive further of a number of interesting avenues of research, such as the influence of changes to the interaction matrix $J$ and extending the model to a more physically relevant one by taking into account the different molecular weights and volumes of the species. In the introduction section, we referred to the experimental situation that raised the domain size question we are addressing here. It is rather difficult to monitor the domain size growth at early stages experimentally. One attempt to capture early stages of the phase separation is to prepare the evaporating films under microgravity conditions and then simulate numerically those scenarios. Alike investigation routes will be studied elsewhere.
	%\cite{microgravity paper?}The direct comparison is very difficult at this stage, due to the limitations in the model and the rather vague definition of the parameters in the model, usually covering a large range of values to study the effect of these. \begin{color}{magenta}More experimental references with interdigitated structures?\end{color}\end{color}

\begin{acknowledgments}
This work was partially funded by the Swedish National Space Agency, grant number 174/19, and the Knut och Alice Wallenbergs Stiftelse, grant number 2016.0059. V.C.E.~K. acknowledges the Lerici foundation for the grant for visiting M.C. at the University of L'Aquila. %\begin{color}{red}Other acknowledgments?\end{color}
\AM{Computational resources from the project SNIC 2019--7--48 are greatly acknowledged.}
\end{acknowledgments}

%\begin{color}{magenta}Check order and style of bibliography!\end{color}


\begin{thebibliography}{0}%
\makeatletter
\providecommand \@ifxundefined [1]{%
 \@ifx{#1\undefined}
}%
\providecommand \@ifnum [1]{%
 \ifnum #1\expandafter \@firstoftwo
 \else \expandafter \@secondoftwo
 \fi
}%
\providecommand \@ifx [1]{%
 \ifx #1\expandafter \@firstoftwo
 \else \expandafter \@secondoftwo
 \fi
}%
\providecommand \natexlab [1]{#1}%
\providecommand \enquote  [1]{``#1''}%
\providecommand \bibnamefont  [1]{#1}%
\providecommand \bibfnamefont [1]{#1}%
\providecommand \citenamefont [1]{#1}%
\providecommand \href@noop [0]{\@secondoftwo}%
\providecommand \href [0]{\begingroup \@sanitize@url \@href}%
\providecommand \@href[1]{\@@startlink{#1}\@@href}%
\providecommand \@@href[1]{\endgroup#1\@@endlink}%
\providecommand \@sanitize@url [0]{\catcode `\\12\catcode `\$12\catcode
  `\&12\catcode `\#12\catcode `\^12\catcode `\_12\catcode `\%12\relax}%
\providecommand \@@startlink[1]{}%
\providecommand \@@endlink[0]{}%
\providecommand \url  [0]{\begingroup\@sanitize@url \@url }%
\providecommand \@url [1]{\endgroup\@href {#1}{\urlprefix }}%
\providecommand \urlprefix  [0]{URL }%
\providecommand \Eprint [0]{\href }%
\providecommand \doibase [0]{https://doi.org/}%
\providecommand \selectlanguage [0]{\@gobble}%
\providecommand \bibinfo  [0]{\@secondoftwo}%
\providecommand \bibfield  [0]{\@secondoftwo}%
\providecommand \translation [1]{[#1]}%
\providecommand \BibitemOpen [0]{}%
\providecommand \bibitemStop [0]{}%
\providecommand \bibitemNoStop [0]{.\EOS\space}%
\providecommand \EOS [0]{\spacefactor3000\relax}%
\providecommand \BibitemShut  [1]{\csname bibitem#1\endcsname}%
\let\auto@bib@innerbib\@empty
%</preamble>
\end{thebibliography}%


\newpage
\begin{thebibliography}{12}

\bibitem{Battaile}
C. Battaile, ``The kinetic Monte Carlo method: Foundation, implementation, and application'', {\sl Comput. Methods Appl. Mech. Engrg.}, \textbf{197}, 3386--3398 (2008).

\bibitem{Rolf}
E. B{\"a}nsch, S. Basting, R. Krahl, ``Numerical simulation of two--phase flows with heat and mass transfer'', {\sl Discrete Continuous Dynamical Systems, Series  A}, \textbf{35}, 6, 2325--2347 (2014).

\bibitem{B1966}
M.\ Blume,
``Theory of the first--order magnetic phase change in UO$_2$'',
{\sl Phys.\ Rev.\/} \textbf{141}, 517 (1966).

\bibitem{BEG1971}
M.\ Blume, V.J.\ Emery, R.B.\ Griffiths,
``Ising model for the $\lambda$ transition and
phase separation in He$^3$--He$^4$ mixtures'',
{\sl Phys.\ Rev.\ A} \textbf{4}, 1071--1077 (1971).

\bibitem{B1994}
A.J.\ Bray,
``Theory of phase--ordering kinetics'',
{\sl Advances in Physics} \textbf{43}, 357--459 (1994).


\bibitem{C1966}
H.W.\ Capel,
``On possibility of first--order phase transitions in Ising systems of
triplet ions with zero--field splitting'',
{\sl Physica} \textbf{32}, 966 (1966).

\bibitem{CGS1997}
E.N.M.\ Cirillo, G.\ Gonnella, S.\ Stramaglia,
``Anisotropic dynamical scaling in a spin model with competing
interactions"
{\sl Physical Review E} \textbf{56}, 5065--5068 (1997).

\bibitem{CGS2005}
E.N.M.\ Cirillo, G.\ Gonnella, G.P.\ Saracco,
``Monte Carlo results for the Ising model with shear"
{\sl Physical Review E} \textbf{72}, 026139 (2005).

\bibitem{CN2013}
E.N.M.\ Cirillo, F.R.\ Nardi,
``Relaxation height in energy landscapes: An application to multiple
metastable states'',
{\sl J.\ Stat.\ Phys.\/} \textbf{150}, 1080--1114 (2013).

\bibitem{CNS2017}
E.N.M.\ Cirillo, F.R.\ Nardi, C.\ Spitoni,
``Sum of exit times in a series of two metastable states'',
{\sl The European Physical Journal Special
Topics} \textbf{226}, 2421--2438 (2017).

\bibitem{CO1996}
E.N.M. Cirillo, E. Olivieri,
``Metastability and nucleation for the Blume--Capel model.
Different mechanisms of transition'',
{\sl J.\ Stat.\ Phys.\/} {\bf 83}, 473--554 (1996).

\bibitem{previous-paper}
E. N. M. Cirillo, M. Colangeli, E. Moons, A. Muntean, S. A. Muntean, J. van Stam,
"A lattice model approach to the morphology formation from ternary mixtures during the evaporation of one component", {\sl
Eur. Phys. J. Special Topics} {\bf 228}, 55--68 (2019).

\bibitem{Col18}
M.\ Colangeli, C.\ Giardin\'{a}, C.\ Giberti, C.\ Vernia, 
``Nonequilibrium two-dimensional Ising model with stationary uphill diffusion'', 
{\sl Phys. Rev. E} \textbf{97}, 030103(R) (2018).

\bibitem{Landau-Binder}
\AM{D. P. Landau, K. Binder, "A Guide to Monte Carlo Simulations in Statistical Physics", Cambridge University Press, 2009.}

  \bibitem{Lowengrub}
J. Cummings, J. S. Lowengrub, B. G. Sumpter, S. M. Wise, R. Kumar, ``Modeling solvent evaporation during thin film formation in phase separating polymer mixtures'', {\sl Soft Matter}, \textbf{14}, 1833--1846 (2018).


\bibitem{DalnokiVeress}
K.~Dalnoki-Veress, J.A.~Forrest, J.R.~Stevens and J.R.~Dutcher.
``Phase separation morphology of spin-coated polymer blend thin films'',
{\sl Physica A}, \textbf{239}, 87-94 (1997).

\bibitem{Olle}
M. Deijfen, O. H{\"a}ggstr{\"o}m, J. Bagley,  ``A stochastic model for competing growth on $\mathrm{R}^d$'', {\sl Markov Processes and Related Fields}, \textbf{10}, 217--248 (2004).

\bibitem{Du}
C. Du, Y. Ji, J. Xue, T. Hou, J. Tang, S.-T. Lee, Y. Li, ``Morphology and performance of polymer solar cell characterized by DPD simulation and graph theory'', Scientific Reports  \textbf{5}, 16854 (2015).

\bibitem{vacancies-paper}
P. Fratzl, O. Penrose,
"Kinetics of spinodal decomposition in the Ising model with vacancy diffusion",
{\sl Phys. Rev. B} \textbf{50}, 3477--3480 (1994).


\bibitem{LL2016}
C.\ Landim, P.\ Lemire,
``Metastability of the two--dimensional Blume--Capel model with
zero chemical potential and small magnetic field'',
{\sl J.\ Stat.\ Phys.\/} \textbf{164},  346--376, (2016).

\bibitem{Lyons1}
B.P. Lyons, N. Clarke, C. Groves, ``The relative importance of domain size, domain purity and domain interfaces to the performance of bulk-heterojunction organic photovoltaics'', {\sl Energy Environ. Sci.}, \textbf{5}, 5, 7657--7663 (2012).

\bibitem{Lyons2}
B.P. Lyons, N. Clarke, C. Groves, ``The quantitative effect of surface wetting layers on the performance of organic bulk heterojunction photovoltaic devices'', {\sl J. Phys. Chem. C}, \textbf{45},  115, 22572--22577 (2011).

\bibitem{BFCLP2007}
M.\ Ib\'a\~nez de Berganza, E.E.\ Ferrero, S.A.\ Cannas,
V.\ Loreto, and A.\ Petri,
``Phase separation of the Potts model in the square lattice'',
{\sl The European Physical Journal Special
Topics} \textbf{143}, 273--275 (2007).

\bibitem{K1972}
K.\ Kawasaki,
``Kinetics of Ising model'',
in C.\ Domb and M.S.\ Green (eds.): Phase Transitions and Critical
Phenomena, Vol.\ 2, Acad. Press, 1972, pp.\ 443--501.

\bibitem{Alison}
L. Meng, Y. Shang, Q. Li,Y. Li, X. Zhan, Z. Shuai, R. G. E. Kimber, A. B. Walker, ``Dynamic Monte Carlo simulation for highly efficient polymer blend photovoltaics'', {\sl J. Phys. Chem. B},  \textbf{114}, 1, 36--41 (2010).

\bibitem{MRRTT1953}
N.\ Metropolis, A.W.\ Rosenbluth, M.N.\ Rosenbluth,
A.H.\ Teller, and E. Teller,
``Equation of state calculations by fast computing machines'',
{\sl J.\ Chem.\ Phys.\/} \textbf{21}, 1087 (1953).

\bibitem{Michels} J. Michels, E. Moons, ``Simulation of surface-directed phase separation in a solution-processed polymer/PCBM blend'', {\sl Macromolecules}, \textbf{46}, 12,  8693 (2013).


\bibitem{Negi}
V. Negi, O. Wodo, J. J. van Franeker, R. A. J. Jansen, P. A. Bobbert, ``Simulating phase separation during spin coating of a polymer-fullerene blend: a joint computational and experimental investigation",  {\sl ACS Appl. Energy Mater.}, \textbf{1}, 2, 725--735 (2018).

\bibitem{Barkema}
\AM{M. E. J. Newman, G. T. Barkema, "Monte Carlo Methods in Statistical Physics", Clarendon Press, Oxford, 2001.}

\bibitem{Onsager1944} L.~Onsager.
\newblock ``Crystal statistics. (i). A two-dimensional model with an order-disorder transition'',
\newblock {\sl Phys. Rev.}, 65:\penalty0 117--149, 1944.

\bibitem{P1952}
R.B.\ Potts,
``Some generalized order--disorder transformations'',
{\sl Proceedings of the Cambridge Philosophical Society}
\textbf{48}, 106--109 (1952).

\bibitem{Harting}
O. J. J. Ronsin, D. J. Jang, H.-J. Egelhaaf, C. J. Brabec, J. Harting, `A phase-field model for the evaporation of thin film mixtures'', {\sl Phys. Chem. Chem. Phys.}, \textbf{22}, 6638--6652 (2020).

\bibitem{Harting2}
\AM{O. J. J. Ronsin, D. J. Jang, H.-J. Egelhaaf, C. J. Brabec, J. Harting, "Phase-field simulation of liquid-vapor equilibrium and evaporation of fluid mixtures",  ACS Appl. Mater. Interfaces \textbf{13}, 55988--56003 (2021).}

  \bibitem{lambda-paper}
    \AM{M. Setta, V. C. E. Kronberg, S. A. Muntean, E. Moons,  J. van Stam, E. N. M. Cirillo, M. Colangeli, A. Muntean,
     "A mesoscopic lattice model for morphology formation in ternary mixtures with evaporation",
       {\sl arXiv}: 2106.01427, (2021).}


\bibitem{Schaefer1}
C. Sch{\" a}fer, J. J. Michels, P. van der Schoot, ``Structuring of thin-film polymer mixtures upon solvent evaporation",  {\sl Macromolecules}, \textbf{49}, 18, 6858--6870 (2016).

\bibitem{Schaefer2}
C. Sch{\" a}fer, S. Paquay, T. C. B. McLeish, ``Morphology formation in binary mixtures upon gradual destabilisation", {\sl Soft Matter}, \textbf{15}, 42, 8450--8458 (2019).


\bibitem{SM1995}
C.\ Site, S.N.\ Majumdar,
``Correlations and coarsening in the $q$--state Potts model'',
{\sl Phys.\ Rev.\ Lett.\/} \textbf{74}, 4321 (1995).

\bibitem{Sprenger}
M. Sprenger, S. Walheim, A.  Budkowski, U.  Steiner,
``Hierarchic structure formation in binary and ternary polymer blends'',
{\sl Interface Science}, \textbf{11}, 225--235 (2003).

\bibitem{Walheim}
S. Walheim, M. B{\" o}ltau, J. Mlynek, G. Krausch, U. Steiner, ``Structure formation via polymer demixing in spin-cast films'', {\sl Macromolecules}, \textbf{30}, 17, 4995--5003 (1997).

\bibitem{Wu1982}
F.Y.\ Wu,
``The Potts model'',
{\sl Review of Modern Physics} \textbf{54}, 235--268 (1982).

\bibitem{Fuwen2018}
F. Zhao, C. Wang, X. Zhan, ``Morphology control in organic solar cells'', {\sl Advanced Energy Materials}, \textbf{8}, 28, 1703147 (2018).


\end{thebibliography}

\end{document}